\documentclass[aps,twocolumn,preprintnumbers,floats,nofootinbib]{revtex4}\def\@cite#1#2{\textsuperscript{[{#1\if@tempswa , #2\fi}]}}
\usepackage{float}
\usepackage{ulem}
\usepackage{graphicx}
\usepackage{amsmath}
\usepackage{bm}
\usepackage{amsfonts}
\usepackage{amssymb}
\usepackage{color}
\usepackage{subfigure}
\usepackage{epsfig}
\usepackage{morefloats}
\usepackage{multirow}
\usepackage{graphicx,booktabs}
\usepackage{mathrsfs}
\usepackage{txfonts}
\usepackage{pifont}
\usepackage{indentfirst}
\usepackage{graphicx,booktabs}
\usepackage{longtable,lscape}
\usepackage[figuresright]{rotating}
\usepackage[colorlinks,citecolor=blue,anchorcolor=red,menucolor=red,linkcolor=red,filecolor=red,urlcolor=blue,frenchlinks=red]{hyperref}
\newcommand{\vrho}{\mbox{\boldmath$\rho$\unboldmath}}
\newcommand{\vlab}{\mbox{\boldmath$\lambda$\unboldmath}}

\begin{document}
	
	\title{Hidden charm pentaquarks and the nature of $P_{c}$ states observed at LHCb}

	\author{ Zhi-Biao Liang$^{1}$, Jun-Jie Liu$^{1}$, Mu-Yang Chen$^{1,4}$~\footnote {E-mail: muyang@hunnu.edu.cn},
		Xian-Hui Zhong$^{1,4}$ ~\footnote {E-mail: zhongxh@hunnu.edu.cn}, Qiang Zhao$^{2,3}$ ~\footnote {E-mail: zhaoq@ihep.ac.cn}}
	\affiliation{ 1) Department of Physics, Hunan Normal University, and Key Laboratory of Low-Dimensional Quantum Structures and Quantum Control of Ministry of Education, Changsha 410081, China }
\affiliation{ 2) Institute of High Energy Physics, Chinese Academy of Sciences, Beijing 100049, China}
\affiliation{ 3) University of Chinese Academy of Sciences, Beijing 100049, China}
	\affiliation{4) Synergetic Innovation Center for Quantum Effects and Applications (SICQEA),
		Hunan Normal University, Changsha 410081, China}

\begin{abstract}

We carry out a unified study of the low-lying $1S$-wave compact states and hadronic
molecules composed of hidden charm pentaquarks $qqqc\bar{c}$ ($q=u,d$) within a semirelativistic potential quark model.
Apart from the linear confinement and one-gluon exchange potential between quarks and/or antiquarks, one-boson exchange potential is also included for baryon and meson clusters within the pentaquark system. We also evaluate the fall-apart decays by combining the obtained spectra within the quark exchange model.
It is found that the $P_c (4312)^+$, $P_c (4440)^+$, and $P_c (4457)^+$ observed by the LHCb Collaboration in 2019 can be well explained by the hadronic
molecules of $[\Sigma_c\bar{D}]_{1/2^-}^{1/2}(4318)$, $[\Sigma_c\bar{D}^*]_{1/2^-}^{1/2}(4437)$, and $[\Sigma_c\bar{D}^*]_{3/2^-}^{1/2}(4458)$, respectively. Meanwhile, $P_c(4380)^+$ reported by LHCb in 2015 may be assigned as the molecule $[\Sigma_c^*\bar{D}]_{3/2^-}^{1/2}(4382)$, except that it turns to be a narrow state other than a broad one shown by the experimental data. Our study shows that the $\sigma$- and $\rho$-meson exchanges are crucial for the formation of $\Sigma_c^{(*)}\bar{D}^{(*)}$ bound states with isospin $I=1/2$. Depending on the potential strength of the $\sigma$ exchange, there may exist very shallow bound states of $\Lambda_c\bar{D}^{(*)}$ with isospin $I=1/2$ and $\Sigma_c^{(*)}\bar{D}^{(*)}$ with isospin $I=3/2$.
Our study may provide useful information for further exploring the hidden-charm pentaquarks in future experiments.
	 	
\end{abstract}
	
	
	\maketitle

\section{Introduction}\label{Fram}

In July 2015, the LHCb collaboration first reported the observations of two hidden charm pentaquark candidates $P_c (4380)^+$ and $P_c (4450)^+$ in the $p J/\psi$ invariant spectrm in  $\Lambda_b\to J/\psi pK^-$~\cite{LHCb:2015yax}. A subsequent higher-statistics analysis resolved the broad structure $P_c (4450)^+$ into two narrow states $P_c (4440)^+$ and $P_c (4457)^+$, and revealed a new state $P_c (4312)^+$~\cite{LHCb:2019kea}. However, their spin-parity quantum
numbers cannot be determined based on the present measurement.
The masses of the newly observed $P_c$ states just lie in the predicted range of early works based on the hadronic molecule picture~\cite{Wu:2012md,Wu:2010jy,Wu:2010vk,Wang:2011rga,Yang:2011wz,Xiao:2013yca}.
The discovery of $P_c$ states at LHCb stimulated great
enthusiasm in the study of hidden-charm pentaquarks from different aspects,
including mass spectra~\cite{Wang:2019ato,He:2019ify,Yamaguchi:2019seo,Liu:2019zvb,Park:2017jbn,Zhu:2019iwm,Shimizu:2019ptd,Ke:2023nra,Ke:2019bkf,Wang:2022ltr,Burns:2019iih,Weng:2019ynv,Wang:2019got,Ali:2019npk,Wang:2025qtm,Chen:2015loa,He:2015cea,Roca:2015dva,Maiani:2015vwa,Lebed:2015tna,Wang:2015epa,
Chen:2015moa,Wu:2017weo,Chen:2019asm,Chen:2019bip,Liu:2019tjn,Xiao:2019aya,
Meng:2019ilv,Cheng:2019obk,Zhang:2019xtu,Mutuk:2019snd,Ruangyoo:2021aoi}, decays~\cite{Xiao:2019mvs,Sakai:2019qph,Wang:2019hyc,Gutsche:2019mkg,Dong:2020nwk,Chen:2020pac,Li:2025ejt,Lin:2017mtz,Wang:2019spc,Deng:2026gqe,Lu:2016nnt,Ortega:2016syt}, productions~\cite{Wang:2025ecf,Bayar:2016ftu,Ling:2021sld,Liu:2021ojf,Shi:2022ipx,Paryev:2022wov,Cao:2023rhu,Liu:2023wfo,Pan:2023hrk,Duan:2024hby,Zhang:2024dkm,Clymton:2026ahm,Guo:2015umn,Liu:2015fea,Wang:2015jsa,Lu:2015fva,Liu:2016dli,Li:2017ghe,Huang:2019jlf,
Guo:2019kdc,Cao:2019kst,Wang:2019krd,Wang:2019dsi,Wu:2019rog,Du:2019pij,Kuang:2020bnk,Xie:2020niw,
Xiao:2020frg,Burns:2022uiv,Du:2021fmf,Nakamura:2021qvy}, and so on.
Except for hadronic molecules~\cite{Wang:2025ecf,Wang:2019ato,He:2019ify,Yamaguchi:2019seo,Liu:2019zvb,Ke:2019bkf,Burns:2019iih,Shimizu:2019ptd,Xiao:2019aya,Guo:2019kdc,Huang:2019jlf,Meng:2019ilv,Liu:2019tjn,Chen:2019asm,
Chen:2015moa,Chen:2015loa,He:2015cea,Roca:2015dva,Chen:2019bip,Zhang:2019xtu}, some other interpretations,
such as compact multiquark states~\cite{Weng:2019ynv,Ruangyoo:2021aoi,Wang:2019got,Ali:2019npk,Zhu:2019iwm,Mutuk:2019snd,Cheng:2019obk,Wu:2017weo,Maiani:2015vwa,Lebed:2015tna,Wang:2015epa} and threshold/kinematical effects~\cite{Guo:2015umn,Liu:2015fea,Bayar:2016ftu,Nakamura:2021qvy,Burns:2022uiv}, were also proposed.

To understand the nature of the $P_c$ states, in particular, whether they favor the
compact pentaquark assignments or the hadronic molecule assignments,
in this work we study these two scenarios of the hidden charm pentaquarks $qqqc\bar{c}$ ($q=u/d$) within a unified quark model framework. For a more comprehensive comparison with the data,
we not only care about the mass spectra of the $1S$-wave states, but also evaluate
their fall-apart decay properties.

To calculate the mass spectrum, we adopt a semi-relativistic hybrid quark potential model
as our previous study of the nucleon and $\Delta$ baryon spectra~\cite{Zhong:2024mnt}.
In this model, besides the short-ranged linear confinement and one-gluon-exchange
(OGE) potentials widely adopted in the literature, we also include the long-ranged one-boson exchange (OBE) potentials
which are crucial for the formation of hadronic molecules. In Ref.~\cite{Zhong:2024mnt},
it is found that the OBE interactions play an important role in the light baryon spectra.
This hybrid quark model including both the OGE and OBE potentials
has been widely applied in the study of the hadron spectra
and hadron-hadron interactions in the literature, e.g., Refs.~\cite{Huang:2015nja,He:2023ucd,Lu:2024dtb,Dai:2003dz,Huang:2005hy,Vijande:2004he,Brauer:1990kt,Zhang:1994pp,Yu:1995ag,Valcarce:2005em,Valcarce:2005rr,Valcarce:2008dr,
Wang:2011rga,Huang:2004sj,Yang:2017qan,Huang:2017dwn}.

Furthermore, to evaluate the fall-apart decays, we adopt a quark-exchange model~\cite{Barnes:2000hu,Barnes:1991em},
which has achieved a good description of the low-energy $S$-wave phase shift for the $I=2$ $\pi\pi$ scattering at the quark level.
In this model, the fall apart decay processes of a multiquark state occur via inner quark rearrangement,
which are assumed to be induced by the short range OGE interactions. Recently, this model has also been extended to study the fall-apart decays of multiquark states in the literature~\cite{Liang:2024met,Liu:2014eka,Wang:2019spc,Xiao:2019spy,Wang:2020prk,Han:2022fup,Liu:2024fnh,Liu:2022hbk,liu:2020eha}, and a lot of inspiring results have been obtained.

Finally, it should be mentioned that there are two significant differences between the compact multiquark state scenario and hadronic molecule picture.
Firstly, for a compact state,
the identical quarks in it should satisfy the requirement of the permutation symmetry.
Under permutation transformation, various color configurations should form a complete linear space.
While for a hadronic molecule composed of two color-singlet hadrons $H_1$ and $H_2$, the quarks between $H_1$ and $H_2$ are not constrained by the permutation symmetry. If permutation symmetry is imposed on molecular states,
except for the color singlet, other color configurations must appear.
When the linear space is complete, the wave functions are equivalent to those of compact states.
Secondly, the differences exist in the interactions. For a compact state, quark-quark interactions
related to the identical quarks have permutation symmetries. While for a loosely
bound molecular state $|H_1H_2\rangle$, for example, if $q_i,q_j\in H_1$ and $q_k\in H_2$, the interactions between
$q_i$ and $q_j$ contributed by both the OGE and OBE are different from the interactions between
$q_i$ and $q_k$ contributed by the OBE only, even $q_j$ and $q_k$ are identical quarks.

This paper is organized as follows. The framework is given in Sec.~\ref{Fram}.
Our results and discussions are presented in Sec.~\ref{sec:results}.
Finally, a summary is given in Sec.~\ref{sec:summary}.

\section{Framework}\label{Fram}
	
In this part, we give a brief review of the potential model for the spectrum calculations,
the quark-exchange model for dealing with the fall-apart decays.

\subsection{Potential model}
	
\subsubsection{Hamiltonian}
	
To investigate the mass spectrum of the $1S$-wave pentaquark states for the $qqqc\bar{c}$ system,
we adopt a semirelativistic constituent quark model. The model Hamiltonian is given by
	\begin{eqnarray}\label{abc}
		H=\sum^{5}_{i}\sqrt{\bm{p_i}^2+m_i^2}+\sum^{5}_{i<j}V_{ij}(r_{ij}),
	\end{eqnarray}
where $m_{i}$ and $\bm{p_i}$ represent the mass and momentum of the $i$-th quark, respectively.
While $V_{ij}$ denotes the effective potentials between the $i$-th and $j$-th quarks with a distance
$r_{ij}=|\bm{r_i}-\bm{r_j}|$. To achieve a more reliable description of the mass spectrum,
in this work the effective potential is taken as a hybrid form
	\begin{eqnarray}
		\label{vij}
		V_{ij}(r_{ij})=V^{OGE}_{ij}(r_{ij})+V^{OBE}_{ij}(r_{ij}),
	\end{eqnarray}
which includes not only includes the potential $V_{ij}^{OGE}(r_{ij})$ widely adopted in the one-gluon-exchange (OGE) potential models,
but also the one-boson-exchange (OBE) potential arising from chiral dynamics.
	
The $V_{ij}^{OGE}(r_{ij})$ potential is adopted as
\begin{eqnarray}\label{eq:conf}
	V_{ij}^{OGE}(r_{ij}) &&= -\frac{3}{16}(\boldsymbol{\lambda}_i^c \cdot \boldsymbol{\lambda}_j^c)\biggl[(b_{ij} r_{ij} + C_{ij})-\frac{4}{3}\frac{\alpha_{ij}}{r_{ij}}\nonumber\\
       && +
	\frac{32\alpha_{ij}\pi}{9m_im_j}\frac{e^{-r_{ij}^2/r_0^2}}{\pi^{3/2}r_0^3}
		(\boldsymbol{S}_i\cdot\boldsymbol{S}_j)\biggr],
\end{eqnarray}
where the first, second, and last terms correspond to the linear confinement part $V^{\mathrm{Conf}}$, color-Coulomb part $V^{\mathrm{Coul}}$, and chromo-magnetic (spin-spin) part $V^{\mathrm{SS}}$, respectively. In the above equation, the parameter $b_{ij}$ denotes the strength of the confinement potential between the $i$-th and $j$-th quarks, while $C_{ij}$ is the corresponding zero point energy. $\bm{\lambda}_{i/j}$ and $\bm{S}_{i/j}$ stand for the color
and spin operators acting on the $i/j$-th quark, respectively. While $\alpha_{ij} $ stands for the effective strong coupling constants, which can be expressed as a product of the individual OGE coupling constants $g_i$ and $g_j$ for the $i$-th and $j$-th quarks,
respectively, i.e., $\alpha_{ij} \equiv g_i g_j$~\cite{Huang:2015nja}.
For $r_0$, we choose a constituent quark mass dependent form as suggested in Ref.~\cite{Silvestre-Brac:1996myf}:
\begin{equation}
	r_0(m_i, m_j) = A\left(\frac{2m_im_j}{m_i+m_j}\right)^{-B},
\end{equation}
where $A$ and $B$ are two free parameters. Finally, it should mentioned that for the quark-quark (antiquark-antiquark) interaction, $\boldsymbol{\lambda}_i^c \cdot \boldsymbol{\lambda}_j^c \equiv \sum_{a=1}^{8} \lambda_i^a \cdot \lambda_j^a$, while for the quark-antiquark interaction  $\boldsymbol{\lambda}_i^c \cdot \boldsymbol{\lambda}_j^c \equiv -\sum_{a=1}^{8} \lambda_i^a \cdot \lambda_j^{a*}$, where $\lambda^{a*}$ is the complex conjugate of the Gell-Mann
matrices $\lambda^{a}$ ($a=1,\cdots,8$) associated with the quark color wave function.

The OBE potentials include the possible contributions from
the pseudoscalar mesons ($\pi$, $K$, $\eta$, and $\eta'$), scalar meson ($\sigma$), and vector mesons ($\rho$, $\omega$, $K^*$, and $\phi$) exchanges, i.e.,  	
	\begin{equation}
	V_{ij}^{OBE}(r_{ij})= \sum_{\chi=\pi,K, \eta,\eta'} V_{\chi}(r_{ij})+ V_{\sigma}(r_{ij})+ \sum_{v=\rho,\omega,K^*,\phi} V_{v}(r_{ij}).
\end{equation}
It should be mentioned that only the center parts of the OBE potentials contributes
to the $1S$-wave states considered in the present work. Their explicit forms of the center parts are given by
\begin{eqnarray}\label{eqution 1}
		\begin{split}	
V_{\chi}^{c}(r_{ij})=&\frac{g_{\chi}^2}{4\pi}\frac{\Lambda^2_{\chi}}{\Lambda^2_{\chi}-m_{\chi}^2}\frac{m_{\chi}^3}{3m_im_j}\left [ Y(m_{\chi }r_{ij})-\frac{\Lambda^3_{\chi}}{m_{\chi}^3}Y(\Lambda_{\chi} r_{ij})  \right ] \\
			&\cdot(\boldsymbol{S} _i\cdot \boldsymbol{S}_j) \mathcal{I}_{\chi},
	\end{split}
\end{eqnarray}
\begin{eqnarray}\label{eqution 2}
		\begin{split}	
&V_{\sigma}^{c}(r_{ij})=-\frac{g_{\sigma}^2}{4\pi}\frac{\Lambda_{\sigma}^2}{\Lambda_{\sigma}^2-m_{\sigma}^2}m_{\sigma}\left [ Y(m_{\sigma }r_{ij})-\frac{\Lambda_{\sigma}}{m_{\sigma }}Y(\Lambda_{\sigma} r_{ij})  \right ]\mathcal{I}_{\sigma},
	\end{split}
\end{eqnarray}
\begin{eqnarray}\label{eqution 3}
\begin{split}
V_{v}^{c}(r_{ij})
		&={}  \frac{g_{v}^{2}}{4\pi}
		\frac{\Lambda_{v}^{2}}
		{\Lambda_{v}^{2}-m_{v}^{2}}\,
		m_{v}\biggl\{\biggl[Y(m_{v}r_{ij})
		-\frac{\Lambda_{v}}{m_{v}}
		Y(\Lambda_{v}r_{ij})\biggr] \\
		&+ \frac{2m_{v}^{2}}{3m_{i}m_{j}} \biggl[Y(m_{v}r_{ij})
		-\Bigl(\frac{\Lambda_{v}}{m_{v}}\Bigr)^{\!3}
		Y(\Lambda_{v}r_{ij})\biggr](\boldsymbol{S} _i\cdot \boldsymbol{S}_j)\biggl\}\mathcal{I}_{v},
	\end{split}
\end{eqnarray}
where $Y(x)$ is the standard Yukawa function $Y(x)=\frac{e^{-x}}{x}$. $m_{\chi}$, $m_{\sigma}$, and $m_{v}$ stand for the masses of the pseudoscalar mesons, scalar $\sigma$ meson, and vector mesons, respectively. $g_{\chi/\sigma/v}$ are the quark-meson-field coupling constants. The $\Lambda$ stands for the cutoff parameters of the meson fields, which characterize the energy scale of spontaneous chiral symmetry breaking.
$\mathcal{I}_{\chi}$ and $\mathcal{I}_{v}$ stand for the flavor operators of the pseudoscalar mesons and vector mesons, respectively.
They are explicitly expressed as follows,
\begin{eqnarray}
	\mathcal{I}_{\pi,\rho} = \sum_{a=1}^{3}
	(\lambda_{i}^{a}\cdot\lambda_{j}^{a}),~
	\mathcal{I}_{K,K^*} = \sum_{a=4}^{7}
	(\lambda_{i}^{a}\cdot\lambda_{j}^{a}),~\mathcal{I}_{\sigma}= \lambda_{i}^{0}\cdot\lambda_{j}^{0},\label{eq:flavor_ops_pi_K_sigma}\\
	\mathcal{I}_{\eta}= \lambda_{i}^{\eta}\cdot\lambda_{j}^{\eta},
\mathcal{I}_{\eta'}= \lambda_{i}^{\eta'}\cdot\lambda_{j}^{\eta'},\mathcal{I}_{\omega}
=\lambda_{i}^{\omega}\cdot\lambda_{j}^{\omega},\mathcal{I}_{\phi}
=\lambda_{i}^{\phi}\cdot\lambda_{j}^{\phi}\label{eq:flavor_ops_eta_eta_omega_phi}.
\end{eqnarray}
The matrices $\lambda^{\eta,\eta',\omega,\phi}$ appearing in Eq.~(\ref{eq:flavor_ops_eta_eta_omega_phi}) are defined as
$\lambda^{\eta}=\lambda^{8}\cos\theta_{P}- \lambda^{0}\sin\theta_{P}$, $\lambda^{\eta'}=\lambda^{8}\sin\theta_{P}+\lambda^{0}\cos\theta_{P}$, $\lambda^{\omega}=\lambda^{8}\sin\theta_{V}+ \lambda^{0}\cos\theta_{V}$, and $\lambda^{\phi}=\lambda^{8}\cos\theta_{V}- \lambda^{0}\sin\theta_{V}$, where $\theta_P$ and $\theta_V$ are the mixing angles for the pseudoscalar and vector mesons in the $SU(3)$ flavor basis, respectively. They are taken as the standard values of the Particle Data Group (PDG), i.e., $\theta_P = -15^{\circ}$ and $\theta_V = 35.3^{\circ}$~\cite{ParticleDataGroup:2024cfk}. The $\lambda^{0}$ is defined as $\lambda^0=\sqrt{\frac{2}{3}}\,\mathbb{I}$, where $\mathbb{I}$ is the $3\times3$ identity matrix. While $\lambda^{a}$ ($a=1,\cdots,8$) are the Gell-Mann matrices associated with the quark flavor wave function.

\begin{table}[htbp]
	\centering
	\caption{Model parameters of the potential model. }
	\label{tab:optimal_params}
	\begin{tabular}{cccccc}
		\hline\hline
		{Parameter} & {Value} & {Parameter} & {Value} & {Parameter} & {Value} \\
		\hline
		$m_{u/d}$\,(GeV)                & 0.322  & $m_c$\,(GeV)                    & 1.474  & $m_{s}$ \,(GeV)             & 0.475 \\
		$g_{u/d}$                       & 0.710  & $g_s$                           & 0.627  & $g_c$              & 0.560\\
		$A $\,(GeV$^{B-1}$)             & 1.435  & $B$                             & 0.800  &  $b_{ij}$\,(GeV$^{2}$)       & 0.124 \\
        $C_{nn}$\,(GeV)                 & $-0.433$  & $C_{ns}$\,(GeV)     & $-0.463$  & $C_{nc}$\,(GeV)             & $-0.222$ \\
        $C_{sc}$\,(GeV)                 & $-0.238$  & $C_{cc}$\,(GeV)   & $-0.100$  & $g_{v}$              & $1.7$ \\
	    $f_{\pi}$\,(MeV)           &  93 & $f_{K/\eta/\eta'}$\,(MeV)  & 113  & $\delta$   & 0.576  \\
		$r_{ij}^{\pi/\rho/\omega}$\,(fm)           &  0.30 & $\Lambda_{\chi/\sigma}$\,(GeV)  & 0.66  & $\Lambda_{v}$\,(GeV)  & 0.85  \\
\hline\hline
	\end{tabular}%
\end{table}

\begin{table}[hptb]
\caption{Masses (MeV) of baryons and mesons and corresponding root mean square radii (fm)
and SHO parameters $\alpha$ (GeV). For the baryons, we define the Jacobi coordinates $\vrho=(\boldsymbol{r}_{1}-\boldsymbol{r}_{2})/\sqrt{2}$ and $\vlab=(\boldsymbol{r}_{1}+\boldsymbol{r}_{2}-2\boldsymbol{r}_{3})/\sqrt{6}$, where $\boldsymbol{r}_{1}$ and $\boldsymbol{r}_{2}$ are the coordinates of the two identical quarks, while $\boldsymbol{r}_{3}$ is coordinate of the third quark.
For the mesons, we define the relative coordinate $\boldsymbol{r}=\boldsymbol{r}_{1}-\boldsymbol{r}_{2}$. Experimental masses are taken from PDG~\cite{ParticleDataGroup:2024cfk}. For the $\Xi_{cc}^{*}$ and $\Omega_{cc}^{(*)}$ baryons, the masses are taken the Lattice QCD predictions of Ref.~\cite{Brown:2014ena}.}
		\label{tab:baryon_massesa}
		\centering
		\begin{tabular}{lcccccccccc}
			\hline\hline
			{State}
			& {$M_{\text{th}}$}
			& {$M_{\text{exp}}$}
			& {$\sqrt{\rho^2}$}   & {$\sqrt{\lambda^2}$}
			& {$\alpha_\rho$}   & {$\alpha_\lambda$} \\
			\hline
			\multicolumn{7}{l}{\textit{Ground light baryons}} \\[2pt]
			$p$                    & 948   & 938        & 0.405 & 0.405 & 0.596 & 0.596 \\
			$\Delta(1232)$         & 1245  & 1235         & 0.525 & 0.525 & 0.460 & 0.460 \\
			$\Lambda$              & 1116  & 1116       & 0.425 & 0.425 & 0.568 & 0.568 \\
			$\Sigma$               & 1207  & 1193       & 0.457 & 0.457 & 0.529 & 0.529 \\
			$\Sigma(1385)$         & 1357  & 1383        & 0.527 & 0.527 & 0.459 & 0.459 \\
			\multicolumn{7}{l}{\textit{Radially excited light states}} \\[2pt]
			$N(1440)1/2^+$              & 1455  & {1440(30)}     & 0.653 & 0.653 & 0.370 & 0.370 \\
			$\Delta(1600)3/2^+$         & 1673  & {1570(70)}     & 0.749 & 0.749 & 0.323 & 0.323 \\
			$\Lambda(1600)1/2^+$        & 1599  & {1600(30)}     & 0.664 & 0.664 & 0.364 & 0.364 \\
			$\Sigma(1660)1/2^+$         & 1664  & {1660(20)}     & 0.691 & 0.691 & 0.350 & 0.350 \\
			$\Sigma(1780)3/2^+$         & 1782  & {1780(50)}     & 0.748 & 0.748 & 0.323 & 0.323 \\
			\hline
			\multicolumn{7}{l}{\textit{Ground singly-charmed baryons}} \\[2pt]
			$\Lambda_{c}$          & 2273  & 2286        & 0.396 & 0.409 & 0.610 & 0.592 \\
			$\Sigma_{c}(2455)$     & 2465  & 2454        & 0.500 & 0.432 & 0.483 & 0.560 \\
			$\Sigma_{c}(2520)$     & 2495  & 2518        & 0.512 & 0.447 & 0.472 & 0.541 \\
			\hline
			\multicolumn{7}{l}{\textit{Ground doubly-charmed baryons}} \\[2pt]
			$\Xi_{cc}$             & 3646  & $3622$        & 0.374 & 0.458 & 0.649 & 0.528 \\
			$\Xi_{cc}^{*}$         & 3678  & $3692$~\cite{Brown:2014ena}  & 0.381 & 0.476 & 0.634 & 0.508 \\
			$\Omega_{cc}$          & 3763  & $3738$~\cite{Brown:2014ena}  & 0.375 & 0.443 & 0.645 & 0.546 \\
			$\Omega_{cc}^{*}$      & 3790  & $3822$~\cite{Brown:2014ena}  & 0.382 & 0.461 & 0.632 & 0.525 \\
\hline\hline
			\multicolumn{7}{l}{\textit{Ground meson states involving in the decays}} \\[2pt]
			{State}
			& {$M_{\text{th}}$}
			& {$M_{\text{exp}}$}
			& \multicolumn{2}{c}{$\sqrt{r^2}$}
			& \multicolumn{2}{c}{$\alpha$} \\
			\hline
			$D$                    & 1831  & 1865      & \multicolumn{2}{c}{0.405} & \multicolumn{2}{c}{0.597} \\
			$D^{*}$                & 1998  & 2007      & \multicolumn{2}{c}{0.482} & \multicolumn{2}{c}{0.501} \\
			$\eta_{c}(1S)$         & 3001  & 2984       & \multicolumn{2}{c}{0.361} & \multicolumn{2}{c}{0.669} \\
			$J/\psi(1S)$           & 3071  & 3097       & \multicolumn{2}{c}{0.396} & \multicolumn{2}{c}{0.611} \\
\hline\hline
\end{tabular}
	\end{table}

\subsubsection{Model parameters}


The model parameters adopted in the present work are collected in Table~\ref{tab:optimal_params}.
About their determinations, some discussions are given as follows.

In the OBE potentials, there are four coupling constants, $g_{\chi}$ ($\chi=\pi,K,\eta, \eta'$), for the pseudoscalar meson changes,
four coupling constants, $g_{v}$ ($v=\rho,\omega, K^*, \phi$), for the vector meson changes, and one coupling constant $g_{\sigma}$ for
the scalar $\sigma$ exchange to be determined.
To be consistent with the previous work of our group~\cite{Zhong:2024mnt},
the coupling constants for the pseudoscalar meson changes are determined by
\begin{equation}
	g_{\chi} = \delta\,\frac{m_u}{f_\chi},
\end{equation}
where $\delta$ and $f_\chi$ are a global parameter accounting for the strength of
the quark-pseudoscalar-meson couplings and the corresponding
meson decay constant, respectively, which are associated with the chiral Lagrangian $\mathcal{L}_{ps}=\frac{\delta}{\sqrt{2}f_{\chi }}\bar{\psi}\gamma^{\mu}\gamma^{5}\psi \partial ^{\mu}\phi_{m}$, where $\psi$ stands for the light quark field with $\psi^{\top}\equiv(u,~ d,~s)$, and $\phi_{m}$ is the matrix representation of the pseudoscalar-meson fields. For the parameter $\delta$, we take the same value, $\delta=0.576$, as that determined by the strong decays of light strange baryon resonances in
Refs.~\cite{Xiao:2013xi,Xiao:2018pwe,Liu:2019wdr}.
For the $\pi$, $K$, and $\eta$ mesons, the decay constants are fixed with
$f_{\pi} = 93$~MeV, $f_{K} = f_{\eta^{(\prime)}}= 113$~MeV. For the
vector mesons, the coupling constants
are fixed with $g_{v}= 1.7$, which are associated
with the effective interaction $\mathcal{L}_{v}=g_{v}\bar{\psi}\gamma_{\mu}V^{\mu}\psi$,
where $V^{\mu}$ is the the matrix representation of the vector-meson fields.
These coupling constants are determined by the vector meson photoproduction processes~\cite{Zhao:1998rt,Zhao:1998fn}.
In this work, we take $g_{\sigma}=g_{\pi}$ as an approximation, which holds in the chiral symmetry limit~\cite{Yu:1995ag,Gell-Mann:1960mvl}.

Furthermore, in the OBE potentials, a series cutoff parameters are introduced. For the
pseudoscalar ($\chi=\pi$, $K$, $\eta$, $\eta'$) and scalar ($\sigma$)
mesons, the cutoff parameters are fixed with
$\Lambda_{\chi}
= \Lambda_{\sigma} = 0.66$~GeV as those determined in Ref.~\cite{Zhong:2024mnt}. While for the
vector mesons ($v=\rho,\omega,K^*,\phi$), to consist with our previous work~\cite{Liu:2026akl} we take $\Lambda_{v} = 0.85$~GeV,
which is determined by a global fitting of the light meson spectrum as listed in Table III of Ref.~\cite{Liu:2026akl} and baryon spectrum as listed in Table~\ref{tab:baryon_massesa} of this work.
The masses of the exchange meosns in the OBE potentials are taken their physical values from the PDG~\cite{ParticleDataGroup:2024cfk}, i.e.,
$m_{\pi} = 135$~MeV, $m_{K} = 498$~MeV, $m_{\eta} = 548$~MeV, $m_{\sigma} = 458$~MeV, $m_{\rho} = 770$~MeV, and $m_{\omega} = 782$~MeV.
The constituent masses for the light $u$ and $d$ quarks in the OBE potentials are taken the same
values as those appearing in the OGE potentials. It should be mentioned that at short distance between light quarks
the masses of proton and $\Lambda_c$ baryon are sensitive to $\pi$, $\rho$, and $\omega$ exchanges,
the divergent behavior at short range will lead to instability of numerical results,
thus, we introduce a cutoff distance $r_{ij}^{\pi/\rho/\omega}=0.30$ fm
for the $\pi$-, $\rho$- and $\omega$-exchange potentials, which are determined by
the measured masses of proton and $\Lambda_c$ baryon.

For the OGE potential sector, the constituent masses
and the strong coupling constants ($g_{i/j}$) are taken the same values as those
adopted in our previous work~\cite{Liu:2026akl}, they are determined by a global fitting the light meson
spectrum as listed in Table III of Ref.~\cite{Liu:2026akl} and baryon spectrum as listed in Table~\ref{tab:baryon_massesa} of this work.
The slop parameters $b_{ij}$ and zero energy parameters $C_{ij}$
of the linear confinement potential together with $A$ and $B$ of the chromo-magnetic interactions
are determined by fitting the ground baryon states and a few observed radially excited
states observed in experiments, which have been listed in Table~\ref{tab:baryon_massesa}.
It should be mentioned that the masses for the $\Xi_{cc}^*$ and $\Omega_{cc}^{(*)}$ are
taken the Lattice QCD predictions of Ref.~\cite{Brown:2014ena}, since they have not been established in experiments.


\subsection{Quark model classification}
	
Generally, the total wave function of a hadron system can be expressed as a direct product of the flavor, spatial, spin, and color parts, i.e.,
	\begin{eqnarray}
		|\Psi \rangle=|\mathrm{flavor}\rangle \otimes |\mathrm{spatial} \rangle \otimes |\mathrm{spin} \rangle \otimes |\mathrm{color} \rangle.
	\end{eqnarray}
In the following, we give a brief introduction for constructing the configurations of the compact pentaquarks and hadronic molecules, respectively.
	
\begin{figure*}[htbp]
 \centering \epsfxsize=14 cm \epsfbox{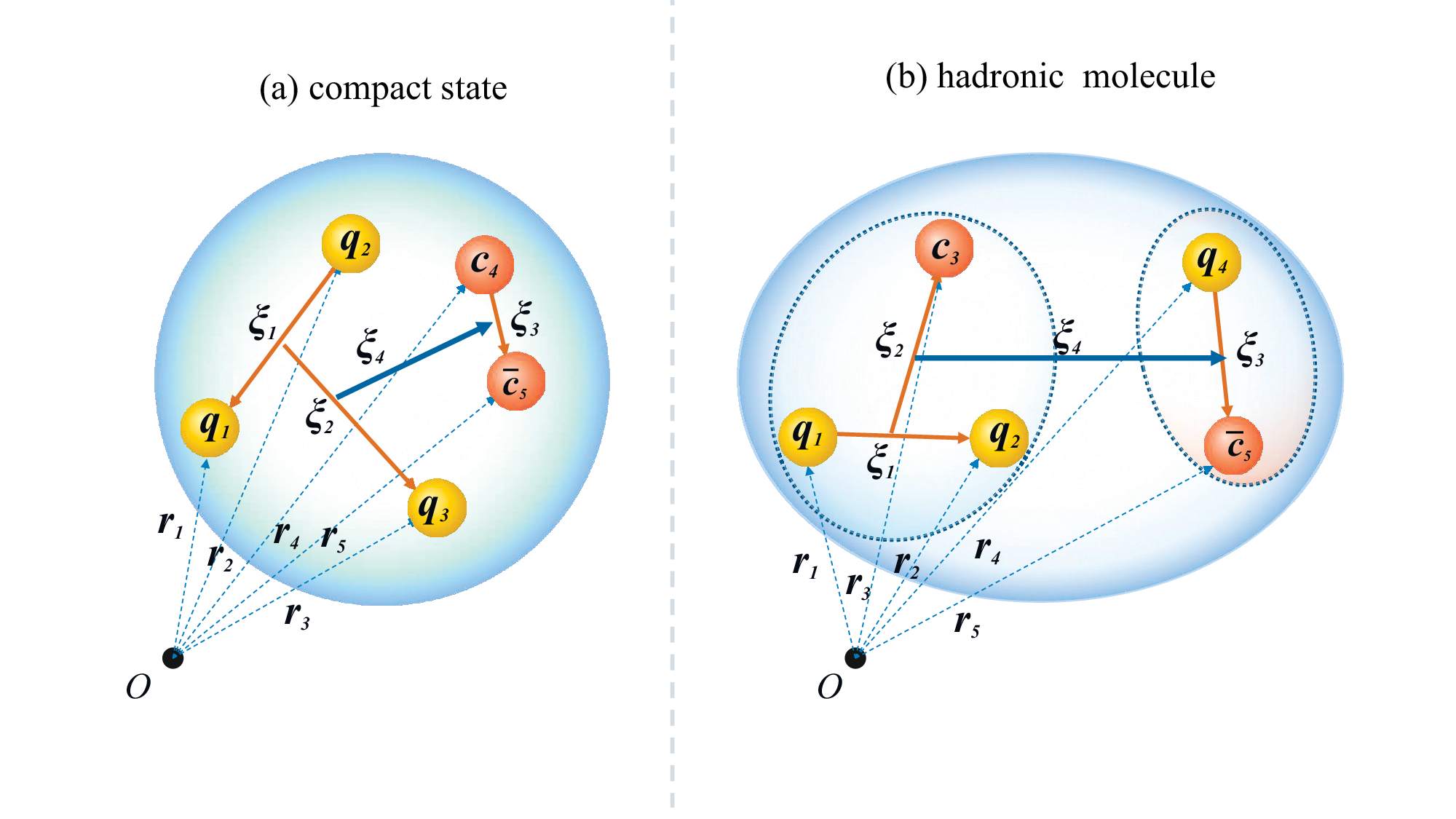}\vspace{-0.5 cm}
 \caption{The illustration of the compact state (a) and hadronic molecule (b) composed of pentaquarks $qqq c\bar{c}$.
}\label{fig:nnncc}
\end{figure*}

\subsubsection{Compact configurations}
	
For a compact $qqqc\bar{c}$ system as shown in Fig~\ref{fig:nnncc} (a), the spatial wave function of the low-lying $1S$-wave states is fully symmetric under the exchange of any two identical light quarks. According to the Pauli principle, the combined flavor-color-spin wave function of the three light quarks ($q_1q_2q_3$) must be fully antisymmetric. We therefore employ the representations of the $S_3$ permutation group to construct the properly antisymmetrized configurations.


First, we consider the flavor space. The flavor wave function of the three light quarks respects the SU(3) symmetry and can be constructed by employing the SU(3) Clebsch--Gordan (C-G) coefficients. The available flavor representations, expressed in terms of Young tableaux, are given by
	\begin{eqnarray}\label{flavor1}
		\begin{aligned}
			F_1=\begin{small}
				\begin{tabular}{lll}
					\cline{1-3}
					\multicolumn{1}{|l|}{1} & \multicolumn{1}{|l|}{2}& \multicolumn{1}{|l|}{3} \\ \cline{1-3}
				\end{tabular}
			\end{small}\otimes(c\bar{c}),\quad
		\end{aligned}
		\begin{aligned}
			F_2=\begin{small}
				\begin{tabular}{ll}
					\cline{1-2}
					\multicolumn{1}{|l|}{1} & \multicolumn{1}{|l|}{2} \\ \cline{1-2}
					\multicolumn{1}{|l|}{3} &                         \\ \cline{1-1}
				\end{tabular}
			\end{small}\otimes(c\bar{c}),
		\end{aligned}\nonumber
	\end{eqnarray}
	
	\begin{eqnarray}\label{flavor2}
		\begin{aligned}
			F_3=\begin{small}
				\begin{tabular}{ll}
					\cline{1-2}
					\multicolumn{1}{|l|}{1} & \multicolumn{1}{|l|}{3} \\ \cline{1-2}
					\multicolumn{1}{|l|}{2} &                         \\ \cline{1-1}
				\end{tabular}
			\end{small}\otimes(c\bar{c}),\quad
		\end{aligned}
		\begin{aligned}
			F_4=
			\begin{small}
				\begin{tabular}{l}
					\cline{1-1}
					\multicolumn{1}{|l|}{1} \\ \cline{1-1}
					\multicolumn{1}{|l|}{2} \\ \cline{1-1}
					\multicolumn{1}{|l|}{3} \\ \cline{1-1}
				\end{tabular}
			\end{small}
			\otimes(c\bar{c}).
		\end{aligned}
	\end{eqnarray}
	Here $F_1$ corresponds to the fully symmetric $\mathbf{10}$ representation (yielding $I=3/2$ states such as $\Delta$-like configurations), $F_2$ and $F_3$ correspond to the mixed-symmetry $\mathbf{8}$ representations (yielding $I=1/2$ states such as $N$-like configurations), and $F_4$ corresponds to the fully antisymmetric $\mathbf{1}$ representation.
	
	In the color space, by using the C-G coefficients of the SU(3) group~\cite{Kaeding:1995vq}, one can construct
	three independent colorless (color-singlet) configurations for the pentaquark system, which can also be found in the literature, e.g., Ref.~\cite{Zhang:2023hmg}.
	These three colorless configurations $C_1$, $C_2$, and $C_3$ can be expressed in the form of
	baryon-meson color structures, i.e.,
	\begin{eqnarray}
		\label{color}
		C_1=\begin{tabular}{|c|}
			\hline
			1 \\ \hline
			2 \\ \hline
			3 \\ \hline
		\end{tabular}_1 (4\bar{5})_1, \quad C_2=\begin{tabular}{|c|c|}
			\hline
			1 & 2\\
			\hline
			3 & \multicolumn{1}{c}{}\\
			\cline{1-1}
		\end{tabular}_8\otimes (4\bar{5})_8, \quad C_3=\begin{tabular}{|c|c|}
			\hline
			1 & 3\\
			\hline
			2 & \multicolumn{1}{c}{}\\
			\cline{1-1}
		\end{tabular}_8\otimes (4\bar{5})_8.\nonumber\\
		~~~~~~~~~~~~~~~~~~~
	\end{eqnarray}
	It is found that the $C_1$ configuration factorizes into a product of two color singlets: a color-singlet triquark and a color-singlet $q\bar{q}$ pair. In contrast, the $C_2$ and $C_3$ configurations involve a color-octet triquark coupled to a color-octet $q\bar{q}$ pair. As will be discussed below, this distinction has profound implications for the internal dynamics and the nature of the resulting pentaquark states.
	
For a pentaquark system, the total spin quantum numbers can take the values $S=5/2,3/2,1/2$. Based on the SU(2) symmetry, one can construct
the spin configurations. Their representations with Young tableaux are given by
	\begin{eqnarray}\label{a82}
		S=\frac{5}{2}: S_1=\begin{tabular}{lllll}
			\cline{1-4}
			\multicolumn{1}{|l|}{1} & \multicolumn{1}{|l|}{2} & \multicolumn{1}{|l|}{3} & \multicolumn{1}{|l|}{4} & 5. ~~~~~~~~~~~~~~~~~~~~~~~~~~~~~~~~~~~\\ \cline{1-4}
		\end{tabular}
	\end{eqnarray}
	\begin{eqnarray}\label{a32}
		\begin{aligned}
			\smallskip\smallskip
			S=\frac{3}{2}:S_2&=\begin{tabular}{llll}
				\cline{1-4}
				\multicolumn{1}{|l|}{1} & \multicolumn{1}{|l|}{2} & \multicolumn{1}{|l|}{3} & \multicolumn{1}{|l|}{4} \\ \cline{1-4}
				5                       &                         &                         &\\
			\end{tabular}, \quad\quad
			S_3=\begin{tabular}{llll}
				\cline{1-3}
				\multicolumn{1}{|l|}{1} & \multicolumn{1}{|l|}{2} & \multicolumn{1}{|l|}{3} & 5 \\ \cline{1-3}
				\multicolumn{1}{|l|}{4} &                         &                         &   \\ \cline{1-1}
			\end{tabular},\\\smallskip\smallskip
			S_4&=\begin{tabular}{llll}
				\cline{1-3}
				\multicolumn{1}{|l|}{1} & \multicolumn{1}{|l|}{3} & \multicolumn{1}{|l|}{4} & 5 \\ \cline{1-3}
				\multicolumn{1}{|l|}{2} &                         &                         &   \\ \cline{1-1}
			\end{tabular}, \quad\quad
			S_5=\begin{tabular}{llll}
				\cline{1-3}
				\multicolumn{1}{|l|}{1} & \multicolumn{1}{|l|}{2} & \multicolumn{1}{|l|}{4} & 5 \\ \cline{1-3}
				\multicolumn{1}{|l|}{3} &                         &                         &   \\ \cline{1-1}
			\end{tabular}.~~~~~~~~~~~~~
		\end{aligned}
	\end{eqnarray}
	\begin{eqnarray}\label{a12}
		\begin{aligned}
			\smallskip\smallskip
			S=\frac{1}{2}:S_6&=\begin{tabular}{lll}
				\cline{1-3}
				\multicolumn{1}{|l|}{1} & \multicolumn{1}{|l|}{2} & \multicolumn{1}{|l|}{3} \\ \cline{1-3}
				\multicolumn{1}{|l|}{4} & 5                       &                         \\ \cline{1-1}
			\end{tabular}, \quad
			S_7=\begin{tabular}{lll}
				\cline{1-3}
				\multicolumn{1}{|l|}{1} & \multicolumn{1}{|l|}{3} & \multicolumn{1}{|l|}{4} \\ \cline{1-3}
				\multicolumn{1}{|l|}{2} & 5                       &                         \\ \cline{1-1}
			\end{tabular}, \quad
			S_8=\begin{tabular}{lll}
				\cline{1-3}
				\multicolumn{1}{|l|}{1} & \multicolumn{1}{|l|}{2} & \multicolumn{1}{|l|}{4} \\ \cline{1-3}
				\multicolumn{1}{|l|}{3} & 5                       &                         \\ \cline{1-1}
			\end{tabular}, \\\smallskip\smallskip
			S_9&=\begin{tabular}{lll}
				\cline{1-2}
				\multicolumn{1}{|l|}{1} & \multicolumn{1}{|l|}{2} & 5 \\ \cline{1-2}
				\multicolumn{1}{|l|}{3} & \multicolumn{1}{|l|}{4} &   \\ \cline{1-2}
			\end{tabular}, \quad
			S_{10}=\begin{tabular}{lll}
				\cline{1-2}
				\multicolumn{1}{|l|}{1} & \multicolumn{1}{|l|}{3} & 5 \\ \cline{1-2}
				\multicolumn{1}{|l|}{2} & \multicolumn{1}{|l|}{4} &   \\ \cline{1-2}
			\end{tabular}.
		\end{aligned}
	\end{eqnarray}
By combining the C-G coefficients of the SU(2) group, one can obtain the explicit spin wave function
$|SS_z\rangle$ corresponding to each Young tableau.

With the flavor, color, and spin configurations at hand,one can further construct the fully coupled configurations in the flavor$\otimes$spin$\otimes$color space. For the low-lying $1S$-wave pentaquark configurations, the spatial wave functions are symmetric under the
exchanges of any two identical quarks, hence, the flavor$\otimes$spin$\otimes$color part must be antisymmetric.
To obtain the properly antisymmetrized coupling configurations in the direct product space of flavor$\otimes$spin$\otimes$color, we employ the C-G coefficients of the permutation group, which are taken from Ref.~\cite{Stancu:1999qr}. The complete set of allowed $1S$-wave configurations for the $I=1/2$ and $I=3/2$ sectors is summarized in Table~\ref{tab:configurations}.
	
	\begin{table}[htbp]
		\centering
		\caption{The $1S$-wave configurations for the compact $qqqc\bar{c}$ system.}
		\label{tab:configurations}
		\footnotesize
		\renewcommand{\arraystretch}{1.22}
		\setlength{\tabcolsep}{2.2pt}
		\begin{tabular}{cccccccccccccccccc}
			\hline
			$I$  $J^P$ & Configuration & Wave function & $(qqq)_c \otimes(c\bar{c})_c$   \\
			\hline
			$\frac{1}{2}\frac{5}{2}^-$
			& $1S_{1/2(5/2^-)}$
			& $\sqrt{\frac{1}{2}}(F_2C_3-F_3C_2)S_1$
			& $\mathbf{8}_c \otimes \mathbf{8}_c$
		 \\
			\hline
			
			$\frac{1}{2}\frac{3}{2}^-$
			& $1S_{1/2(3/2^-)_1}$
			& $\sqrt{\frac{1}{2}}(F_2C_3-F_3C_2)S_2$
			& $\mathbf{8}_c \otimes \mathbf{8}_c$
			 \\
			
			& $1S_{1/2(3/2^-)_2}$
			& $\sqrt{\frac{1}{2}}(F_2C_3-F_3C_2)S_3$
			& $\mathbf{8}_c \otimes \mathbf{8}_c$
			 \\
			
			& $1S_{1/2(3/2^-)_3}$
			& $\sqrt{\frac{1}{2}}(F_2S_5+F_3S_4)C_1$
			& $\mathbf{1}_c \otimes \mathbf{1}_c$
		 \\
			
			& $1S_{1/2(3/2^-)_4}$
			& $\frac{1}{2}F_2(-C_2S_4-C_3S_5)$
			& $\mathbf{8}_c \otimes \mathbf{8}_c$\\
           &&$-\frac{1}{2}F_3(C_2S_5-C_3S_4)$&\\
			\hline
			
			$\frac{1}{2}\frac{1}{2}^-$
			& $1S_{1/2(1/2^-)_1}$
			& $\sqrt{\frac{1}{2}}(F_2C_3-F_3C_2)S_6$
			& $\mathbf{8}_c \otimes \mathbf{8}_c$
			 \\
			
			& $1S_{1/2(1/2^-)_2}$
			& $\sqrt{\frac{1}{2}}(F_2S_8+F_3S_7)C_1$
			& $\mathbf{1}_c \otimes \mathbf{1}_c$
			 \\
			
			& $1S_{1/2(1/2^-)_3}$
			& $\sqrt{\frac{1}{2}}(F_2S_9+F_3S_{10})C_1$
			& $\mathbf{1}_c \otimes \mathbf{1}_c$
			 \\
			
			& $1S_{1/2(1/2^-)_4}$
			& $\frac{1}{2}F_2(-C_2S_7-C_3S_8)$
			& $\mathbf{8}_c \otimes \mathbf{8}_c$
			 \\
			&&$-\frac{1}{2}F_3(C_2S_8-C_3S_7)$&\\
			& $1S_{1/2(1/2^-)_5}$
			& $\frac{1}{2}F_2(-C_2S_{10}-C_3S_9)$
			& $\mathbf{8}_c \otimes \mathbf{8}_c$\\
            &&$-\frac{1}{2}F_3(C_2S_9-C_3S_{10})$&\\
			\hline

			$\frac{3}{2}\frac{5}{2}^-$
			& $1S_{3/2(5/2^-)}$
			& $F_1C_1S_1$
			& $\mathbf{1}_c \otimes \mathbf{1}_c$
			 \\
			\hline

			$\frac{3}{2}\frac{3}{2}^-$
			& $1S_{3/2(3/2^-)_1}$
			& $F_1C_1S_2$
			& $\mathbf{1}_c \otimes \mathbf{1}_c$
			 \\

			& $1S_{3/2(3/2^-)_2}$
			& $F_1C_1S_3$
			& $\mathbf{1}_c \otimes \mathbf{1}_c$
			 \\

			& $1S_{3/2(3/2^-)_3}$
			& $\sqrt{\frac{1}{2}}F_1(C_2S_4-C_3S_5)$
			& $\mathbf{8}_c \otimes \mathbf{8}_c$
			 \\
			\hline

			$\frac{3}{2}\frac{1}{2}^-$
			& $1S_{3/2(1/2^-)_1}$
			& $F_1C_1S_6$
			& $\mathbf{1}_c \otimes \mathbf{1}_c$
			 \\

			& $1S_{3/2(1/2^-)_2}$
			& $\sqrt{\frac{1}{2}}F_1(C_2S_7-C_3S_8)$
			& $\mathbf{8}_c \otimes \mathbf{8}_c$
		 \\

			& $1S_{3/2(1/2^-)_3}$
			& $\sqrt{\frac{1}{2}}F_1(C_2S_{10}-C_3S_9)$
			& $\mathbf{8}_c \otimes \mathbf{8}_c$
			 \\
			\hline
		\end{tabular}
	\end{table}

	\begin{table}[tbp]
		\centering
		\caption{The $1S$-wave molecular configurations for the $\Lambda_c \bar{D}^{(*)}$,
and $\Sigma_c^{(*)}\bar{D}^{(*)}$ systems.}
		\label{tab:molecular_states}
		\small
		\renewcommand{\arraystretch}{1.18}
		\setlength{\tabcolsep}{2.2pt}
		\begin{tabular}{ccccccccccccc}
			\hline
			Isospin & Hadron pair & Configuration  \\
			\hline
			$1/2$ & $\Lambda_c \bar{D}$       & $[\Lambda_c \bar{D}]^{1/2}_{1/2^-}$ \\
			$1/2$ & $\Lambda_c \bar{D}^*$     & $[\Lambda_c \bar{D}^*]^{1/2}_{1/2^-, 3/2^-}$  \\
			$1/2$ & $\Sigma_c \bar{D}$        & $[\Sigma_c \bar{D}]^{1/2}_{1/2^-}$  \\
			$1/2$ & $\Sigma_c^* \bar{D}$      & $[\Sigma_c^* \bar{D}]^{1/2}_{3/2^-}$  \\
			$1/2$ & $\Sigma_c \bar{D}^*$      & $[\Sigma_c \bar{D}^*]^{1/2}_{1/2^-, 3/2^-}$  \\
			$1/2$ & $\Sigma_c^* \bar{D}^*$    & $[\Sigma_c^* \bar{D}^*]^{1/2}_{1/2^-, 3/2^-, 5/2^-}$ \\
			\hline
			$3/2$ & $\Sigma_c \bar{D}$        & $[\Sigma_c \bar{D}]^{3/2}_{1/2^-}$  \\
			$3/2$ & $\Sigma_c^* \bar{D}$      & $[\Sigma_c^* \bar{D}]^{3/2}_{3/2^-}$  \\
			$3/2$ & $\Sigma_c \bar{D}^*$      & $[\Sigma_c \bar{D}^*]^{3/2}_{1/2^-, 3/2^-}$  \\
			$3/2$ & $\Sigma_c^* \bar{D}^*$    & $[\Sigma_c^* \bar{D}^*]^{3/2}_{1/2^-, 3/2^-, 5/2^-}$  \\
			\hline
		\end{tabular}
	\end{table}


\subsubsection{Molecular configurations}

For a loose hadronic molecule system $(qqc)(q\bar{c})$ as shown in Fig~\ref{fig:nnncc} (b), the two constituent hadrons $H_1(qqc)$ and $H_2(q\bar{c})$ are considered as two spatially separated color-singlet clusters. We therefore focus on the $H_1$-$H_2$ interactions and do not impose the Pauli antisymmetry condition between light quarks belonging to different clusters.

In color space, both the $H_1$ and $H_2$ subsystems are color singlets in the molecular picture. Thus, only the color-singlet structure ($\mathbf{1}_c \otimes \mathbf{1}_c$) contributes. In this case, the short-range color-Coulomb and chromomagnetic interactions vanish between the two clusters, and the one-boson-exchange (OBE) potentials provide the binding mechanism.

In the isospin space, the flavor wave functions with isospin quantum numbers $II_z$ are couplings of isospin eigenstates  $|I_1I_{1z}\rangle_{H_1}$ and $|I_2I_{2z}\rangle_{H_2}$ of $H_1$ and $H_2$, i.e.,
	\begin{eqnarray}
		|II_z\rangle_{H_1H_2}=\sum_{I_{1z},I_{2z}}C^{II_z}_{I_1I_{1z};I_2I_{2z}}|I_1I_{1z}\rangle_{H_1}|I_2I_{2z}\rangle_{H_2},
	\end{eqnarray}
where $C^{II_z}_{I_1I_{1z};I_2I_{2z}}$ stands for the C-G coefficients. For the $\Lambda_cD^{(*)}$ system, the isospin numbers can take $I=1/2$; while for the $\Sigma_c^{(*)}D^{(*)}$ system,
the isospin can take $I=1/2$ and $I=3/2$.
	
Similarly, the spin wave functions of the $1S$-wave states with spin quantum numbers $JJ_z$ are couplings of angular momentum eigenstates $|j_1j_{1z}\rangle_{H_1}$ and $|j_2j_{2z}\rangle_{H_2}$ of $H_1$ and $H_2$, i.e.,
	\begin{eqnarray}
		|JJ_z\rangle_{H_1H_2}=\sum_{j_{1z},j_{2z}}C^{JJ_z}_{j_1j_{1z};j_2j_{2z}}|j_1j_{1z}\rangle_{H_1}|j_2j_{2z}\rangle_{H_2}.
	\end{eqnarray}

Finally, combining the flavor and spin wave functions, one obtains the $1S$-wave molecular configurations for the $\Lambda_c \bar{D}^{(*)}$,
and $\Sigma_c^{(*)}\bar{D}^{(*)}$ systems considered in the present work, which are summarized in Table~\ref{tab:molecular_states}.
It should be mentioned that these molecular configurations are orthogonal to those compact configurations
given in Table~\ref{tab:configurations}.

	
		
\subsection{Numerical method}
	
To solve the five-body Schr\"{o}dinger equation, we adopt the explicitly correlated Gaussian (ECG) method~\cite{Mitroy:2013eom,Varga:1995dm}.
The ECG method is a variational approach widely used for quantum few-body problems
in molecular, atomic, nuclear, and hadronic physics. In this framework, the spatial
wave function is expanded in a set of correlated Gaussian basis functions.

For a compact pentaquark configurations composed of $q_1q_2q_3c_4\bar{c}_5$ as shown in Fig~\ref{fig:nnncc} (a),
the basis functions adopt a correlated Gaussian form
\begin{eqnarray}\label{aaa}	
\psi(\bm{r}_{1},\bm{r}_{2},\bm{r}_{3},\bm{r}_{4},\bm{r}_{5})=\exp\left(-\sum\limits_{i<j}^5a_{ij}\bm{r}_{ij}^2\right),
\end{eqnarray}
where $a_{ij}$ are variational parameters. According to the permutation symmetry of the three
identical light quarks ($q_1q_2q_3$), the ten variational parameters $a_{ij}$
are reduced to four independent ones: $a_{12}=a_{13}=a_{23}\equiv a$, $a_{14}=a_{24}=a_{34}\equiv b$, $a_{15}=a_{25}=a_{35}\equiv c$,
and $a_{45}\equiv d$. In the calculations, one should select a set of Jacobi coordinates $\bm{\xi}=\left(\bm{\xi}_1, \bm{\xi}_2, \bm{\xi}_3,\bm{\xi}_4\right)$,
such as
		\begin{eqnarray}\label{bbb}
			\begin{aligned}
				\bm{\xi_1} & = \bm{r_1}-\bm{r_2},                                                                                                        \\
				\bm{\xi_2} & = \bm{r_3}-\frac{m_1\bm{r_1}+m_2\bm{r_2}}{m_1+m_2},                                                                 \\
				\bm{\xi_3} & = \bm{r_4}-\bm{r_5},                                                                                                        \\
				\bm{\xi_4} & = \frac{m_1\bm{r_1}+m_2\bm{r_2}+m_3\bm{r_3}}{m_1+m_2+m_3}-\frac{m_4\bm{r_4}+m_5\bm{r_5}}{m_4+m_5}.
			\end{aligned}
		\end{eqnarray}
In terms of the Jacobi coordinates, the general correlated Gaussian basis function
of Eq.~(\ref{aaa}) is transformed into
\begin{eqnarray}
G(\bm{\xi},\mathbb{A})=\exp(-\bm{\xi}^T\mathbb{A}\bm{\xi}),
\end{eqnarray}
where $\mathbb{A}$ is a $4\times4$ positive-definite nondiagonal matrix determined by the variational
parameters.

For a molecular configuration composed of $(q_1q_2c_3)(q_4\bar{c}_5)$ as shown in Fig~\ref{fig:nnncc} (b),
with the Jacobi coordinates defined in Eq.~(\ref{bbb}) the correlated Gaussian basis functions is taken the following form
\begin{eqnarray}\label{aaa}	
\psi_{\rm mol}(\bm{\xi})&=&\exp\left(-a\bm{\xi}_1^2-b\bm{\xi}_2^2\right)
\exp\left(-c\bm{\xi}_3^2\right)\exp\left(-d\bm{\xi}_4^2\right),\nonumber\\
&=& G(\bm{\xi},\mathbb{A})
\end{eqnarray}
where $\bm{\xi}_1$ and $\bm{\xi}_2$ describe the relative motions of quarks in
the baryon cluster ($q_1q_2c_3$), $\bm{\xi}_3$ describes the relative motion
between the quark and antiquark in the
meson cluster $(n_4\bar{c}_5)$, and $\bm{\xi}_4$ describes relative
motion between the two color-singlet clusters. In this case, $\mathbb{A}$ is a
diagonal matrix $\mathbb{A}=\mathrm{diag}(a,b,c,d)$.

The spatial part of the trial wave function for both compact and molecular configurations
can be expanded with the Gaussian functions, i.e.,
\begin{eqnarray}\label{spatial}
\Psi(\bm{\xi},\mathbb{A})=\sum\limits_{k=1}^N c_k G(\bm{\xi},\mathbb{A}_k),
\end{eqnarray}
where $N$ is the number of Gaussian basis functions and $c_k$ are the linear
expansion coefficients. The accuracy of the trial wave function depends on both
the number $N$ and the nonlinear parameter matrices $\mathbb{A}_k$. In our
calculations, following the method of Ref.~\cite{Hiyama:2003cu}, we let the
variational parameters form a geometric progression. For instance,
for a variational parameter $a$, we take
		\begin{eqnarray}
			a_i=\frac{1}{2\left(a_1q^{i-1}\right)^2}\quad\left(i=1,\cdots,n_{max}\right).
		\end{eqnarray}
The Gaussian size parameters $\{a_1,q,n_{max}\}$ are determined through the variational principle.
It should be mentioned that the final results should be independent on the choice of these parameters what we choose.

For a given configuration, the Hamiltonian matrix elements are calculated as
		\begin{eqnarray}
				H_{kk'}=\langle \psi_{FCS}G(\bm{\xi},\mathbb{A}_k) | H |\psi_{FCS}G(\bm{\xi},\mathbb{A}_{k'}) \rangle,
			\end{eqnarray}
where $\psi_{FCS}$ denotes the flavor-spin-color wave function.
 The eigenenergy $E$ and the expansion coefficients
$\{c_k\}$ are then obtained by solving the generalized matrix eigenvalue problem,
\begin{eqnarray}\label{Massa}
\sum_{k'=1}^{N}(H_{kk'}-EN_{kk'})c_{k'}=0,
\end{eqnarray}
where $N_{kk'}=\langle G(\bm{\xi},\mathbb{A}_k)|G(\bm{\xi},\mathbb{A}_{k'})\rangle$ is the overlap matrix.
In the present calculations, stable and converged solutions are obtained with the
basis size $N=n_{max}^a\times n_{max}^b\times n_{max}^c\times n_{max}^d=6\times6\times6\times6=1296$.

\subsection{Fall-apart decay}
	
Once the mass spectrum is obtained, a key question is whether the predicted pentaquark states are stable against dissociation into conventional baryon-meson pairs. For those states lying above the relevant two-body thresholds, we evaluate their fall-apart decay properties within the framework of a quark-exchange model~\cite{Barnes:2000hu,Barnes:1991em}.
In this model, the OGE potential serves as the driving force for the fall-apart decays of the multiquark state via
quark rearrangement into color-singlet hadron pairs. The decay amplitude $\mathcal{M}(A\to BC)$ is
	given by
	\begin{eqnarray}
		\mathcal{M}(A\to BC)=-\sqrt{(2\pi)^3}\sqrt{8M_AE_BE_C}\left\langle BC |\sum_{i\in B,j\in C} V_{ij}^{OGE}| A \right\rangle,
	\end{eqnarray}
where $A$ denotes the initial multiquark state, and $BC$ denotes the final baryon-meson pair.
$V_{ij}^{OGE}$ is the OGE potential given by Eq.~(\ref{vij}).
	$M_A$ is the mass of the initial state, while $E_B$ and $E_C$ are the energies of the final states $B$ and $C$ in the rest frame of the initial hadron, respectively.
The partial decay width for the $A\to BC$ process is then given by
	\begin{eqnarray}
		\Gamma=\frac{1}{2J_A+1}\frac{|\bm{q}|}{8\pi M_A^2}\left|\mathcal{M}(A\to BC)\right|^2,
	\end{eqnarray}
where $\bm{q}$ is the three-momentum of the final state $B$ (or $C$) in the rest frame of the initial hadron. This phenomenological model has achieved a satisfactory description of the low-energy $S$-wave phase shift for the $I=2$ $\pi\pi$
scattering at the quark level~\cite{Barnes:2000hu,Barnes:1991em}. In recent years, this model
has been extensively applied to study the fall-apart decays of various multiquark states~\cite{Wang:2019spc,Xiao:2019spy,Wang:2020prk,Han:2022fup,Liu:2024fnh,Liu:2022hbk,liu:2020eha,Liang:2024met}, yielding a number of valuable predictions.
	
In the present work, the masses and wave functions of the initial pentaquark states are taken from the numerical results of our potential model calculations. For the final-state hadrons $B$ and $C$, for simplicity, the wave functions are approximated by a single
harmonic oscillator (SHO) form. Their SHO parameters are determined by fitting the root mean square radii, which are obtained from our potential model calculations with the same Hamiltonian given in Eq. (\ref{abc}). Our determined root mean square (RMS) radii and SHO parameters for the
final meson and baryon states are collected in Table~\ref{tab:baryon_massesa}.
For the well-established hadron states in the final states, the masses are taken from the PDG averaged
values~\cite{ParticleDataGroup:2024cfk}, which are collected Table~\ref{tab:baryon_massesa}. For the initial pentaquark states, the masses are taken from our potential model predictions.

\begin{table}[hptb]
\caption{Predicted $1S$-wave compact pentaquark states composed of $qqqc\bar{c}$.
The components of different configurations for each states are also given.}
			\label{tab:mass-spectrum}
			\centering
			\renewcommand{\arraystretch}{1.20}
			\begin{tabular}{ccccccccccccc}
				\hline
				{$I(J^{P})$} & {Mass~(MeV)} &
				\multicolumn{3}{c}{Component} \\
				\hline
				& &  \multicolumn{3}{c}{$1S_{1/2(5/2^-)}$} \\
				\cmidrule(lr){3-5}
				{$\dfrac{1}{2}\!\left(\dfrac{5}{2}^{-}\right)$}
				& 4700 & \multicolumn{3}{c}{$100\%$} \\[4pt]
				\hline
				
				&&  {$1S_{1/2(3/2^-)_1}$} & {$1S_{1/2(3/2^-)_2}$} & {$1S_{1/2(3/2^-)_4}$} \\
				\cmidrule(lr){3-5}
				\multirow{3}{*}{$\dfrac{1}{2}\!\left(\dfrac{3}{2}^{-}\right)$}
				& 4630 & $14\%$ &  $\sim 0$ & $86\%$ \\
				& 4669 & $15\%$ & $80\%$ &  $5\%$ \\
				& 4686 & $71\%$ & $20\%$ &  $9\%$ \\[2pt]
				\hline
				
				& & {$1S_{1/2(1/2^-)_1}$} & {$1S_{1/2(1/2^-)_4}$} & {$1S_{1/2(1/2^-)_5}$} \\
				\cmidrule(lr){3-5}
				\multirow{3}{*}{$\dfrac{1}{2}\!\left(\dfrac{1}{2}^{-}\right)$}
				& 4613 & $18\%$ &  0 & $82\%$ \\
				& 4621 & $15\%$ & $82\%$ &  $3\%$ \\
				& 4661 & $67\%$ & $18\%$ & $15\%$ \\
				\hline
				
				& & \multicolumn{3}{c}{$1S_{3/2(3/2^-)_3}$} \\
				\cmidrule(lr){3-5}
				{$\dfrac{3}{2}\!\left(\dfrac{3}{2}^{-}\right)$}
				& 4781 & \multicolumn{3}{c}{$100\%$} \\[4pt]
				\hline
				
				& & {$1S_{3/2(1/2^-)_2}$} & {$1S_{3/2(1/2^-)_3}$} & {} \\
				\cmidrule(lr){3-4}
				\multirow{2}{*}{$\dfrac{3}{2}\!\left(\dfrac{1}{2}^{-}\right)$}
				& 4765 &  $6\%$ & $94\%$ &  \\
				& 4799 & $94\%$ &  $6\%$ &  \\
				\hline
			\end{tabular}
		\end{table}

		\begin{table*}[hptb]
			\caption{Average contributions of each term in the Hamiltonian
				(MeV), and the root-mean-square radii (fm) between different quark pairs, $R_{ij}=\sqrt{r_{ij}^2}$, for the $1S$-wave compact pentaquark states composed of $qqqc\bar{c}$. }
			\label{tab:hamiltonian}
			\centering
			\renewcommand{\arraystretch}{1.15}
			\begin{tabular}{ccccccccccccccccccccccccc}
				\hline\hline
				{$I(J^{P})$} &
				{State} &
				{$\langle T\rangle$} &
				{$\langle V^{\mathrm{Conf}}\rangle$} ~~&
				{$\langle V^{\mathrm{Coul}}\rangle$}~~ &
				{$\langle V^{\mathrm{SS}}\rangle$} ~~&
				{$\langle V_{\pi}\rangle$} ~~&
				{$\langle V_{\sigma}\rangle$}~~ &
				{$\langle V_{\eta}\rangle$} ~~&
				{$\langle V_{\eta^{\prime}}\rangle$} ~~&
				{$\langle V_{\rho}\rangle$} ~~&
				{$\langle V_{\omega}\rangle$}~~
                & $R_{12/13/23}$ & $R_{14/24/34}$& $R_{15/25/35}$ & $R_{45}$ \\
				\hline
				{$\frac{1}{2}(\frac{5}{2}^{-})$}
				&$P_c^N(4700)\frac{5}{2}^-$ & 4898 & 363 & -559 &  27.6
				&  18.3 & $-44.7$ & $-2.9$ & $-0.7$ & $-24.0$ &  23.6 &0.787&0.694& 0.694& 0.663\\
				\hline
				
				\multirow{3}{*}{$\frac{1}{2}(\frac{3}{2}^{-})$}
				& $P_c^N(4630)\frac{3}{2}^-$  & 4969 &  321 & $-585$ &  $-14.6$
				& $-14.9$ & $-47.9$ &  2.5 &  0.6 & $-37.0$ &  36.1&0.749&0.670& 0.665& 0.644 \\
				& $P_c^N(4669)\frac{3}{2}^-$ & 4929 &  344 & $-570$ &   $-1.6$
				&  17.2 & $-45.8$ & $-2.8$ & $-0.7$ & $-25.1$ &  24.7& 0.773 & 0.678 & 0.684 & 0.651 \\
				& $P_c^N(4686)\frac{3}{2}^-$ & 4912 & 355 & $-564$ &   16.6
				&  15.1 & $-45.2$ & $-2.4$ & $-0.6$ & $-25.4$ &  24.9 & 0.780 & 0.689 & 0.689 & 0.660\\
				\hline
				
				\multirow{3}{*}{$\frac{1}{2}(\frac{1}{2}^{-})$}
				& $P_c^N(4613)\frac{1}{2}^-$ & 4987 &  311 & $-591$ & $-33.8$
				& $-13.4$ & $-48.4$ &  2.2 &  0.6 & $-37.1$ &  36.2& 0.742 & 0.661 & 0.660 & 0.636 \\
				& $P_c^N(4621)\frac{1}{2}^-$ & 4978 &  316 & $-588$ & $-24.1$
				& $-14.7$ & $-48.2$ &  2.5 &  0.6 & $-37.2$ &  36.3 & 0.745 & 0.666 & 0.662 & 0.641\\
				& $P_c^N(4661)\frac{1}{2}^-$ & 4938 &  339 & $-573$ & $-0.9$
				&   6.3 & $-46.3$ & $-1.0$ & $-0.2$ & $-28.9$ &  28.3 & 0.767 & 0.676 & 0.681 & 0.651\\
				\hline
				
				{$\frac{3}{2}(\frac{3}{2}^{-})$}
				& $P_c^{\Delta}(4781)\frac{3}{2}^-$ & 4809 & 425 & $-524$ &  36.9
				&  14.8 & $-39.1$ &  2.2 &  0.4 &  28.0 &  27.3 & 0.855 & 0.731 & 0.728 & 0.676\\
				\hline
				
				\multirow{2}{*}{$\frac{3}{2}(\frac{1}{2}^{-})$}
				& $P_c^{\Delta}(4765)\frac{1}{2}^-$ & 4824 & 414 & $-529$ &  20.6
				&  15.1 & $-39.7$ &  2.2 &  0.5 &  28.7 &  27.9 & 0.847 & 0.721 & 0.724 & 0.670\\
				& $P_c^{\Delta}(4799)\frac{1}{2}^-$ & 4793 & 437 & $-517$ &  53.6
				&  14.5 & $-38.5$ &  2.1 &  0.4 &  27.4 &  26.7 & 0.864 & 0.736 & 0.739 & 0.686\\
				\hline\hline
			\end{tabular}
		\end{table*}

	\begin{table*}[hptb]
		\centering
		\caption[Fall-apart decay widths of compact pentaquark states.]{
			Partial fall-apart decay widths $\Gamma_i$ (MeV) of the compact pentaquark states.}
		\label{tab:compact-pentaquark-new}
		\begin{tabular}{cccccccccccc}
			\hline\hline
			State & ~~~~$\Gamma_{p\eta_{c}}$~~~~ & ~~~~$\Gamma_{pJ/\psi}$~~~~ & ~~~~$\Gamma_{\Delta\eta_{c}}$~~~~ & ~~~~$\Gamma_{\Delta J/\psi}$ ~~~~&~~~~ $\Gamma_{\Lambda_{c}\bar{D}}$ ~~~~& ~~~~$\Gamma_{\Lambda_{c}\bar{D}^{*}}$~~~~ & ~~~~$\Gamma_{\Sigma_{c}\bar{D}}$ ~~~~& ~~~~$\Gamma_{\Sigma_{c}\bar{D}^*}$ ~~~~& ~~~~$\Gamma_{\Sigma_{c}^{*}\bar{D}}$~~~~ & ~~~~$\Gamma_{\Sigma_{c}^{*}\bar{D}^{*}}$ ~~~~& $\Gamma_{sum}$ \\
			\hline
			$P_c^{N}(4700)\frac{5}{2}^{-}$ & $\cdots$ & $\cdots$ & $\cdots$ & $\cdots$ & $\cdots$ & $\cdots$ & $\cdots$& $\cdots$ & $\cdots$ & 4.48 & 4.48 \\
			\hline
			$P_c^{N}(4630)\frac{3}{2}^{-}$ &$\cdots$ & 0.33 & $\cdots$ & $\cdots$ & $\cdots$ & ${<}0.01$ & $-$ & 0.95 & 0.83 & 0.31 & 2.43 \\
			$P_c^{N}(4669)\frac{3}{2}^{-}$ & $\cdots$ & 0.10 & $\cdots$& $\cdots$& $\cdots$ & 1.01 & $\cdots$ & 4.26 & 0.24 & ${<}0.01$ & 5.63 \\
			$P_c^{N}(4686)\frac{3}{2}^{-}$ & $\cdots$ & 0.12 &$\cdots$ & $\cdots$ & $\cdots$& 1.27 & $\cdots$ & 1.00 & 1.74 & 0.99 & 5.12 \\
			\hline
			$P_c^{N}(4613)\frac{1}{2}^{-}$ & 1.73 & 0.30 & $\cdots$ &$\cdots$ & 0.10 & 0.06 & 0.37 & 0.85 & $\cdots$ & 2.31 & 5.72 \\
			$P_c^{N}(4621)\frac{1}{2}^{-}$ & 0.05 & 0.71 &$\cdots$ & $\cdots$ & 0.32 & 0.22 & 0.02 & 2.32 & $\cdots$ & 6.35 & 9.99 \\
			$P_c^{N}(4661)\frac{1}{2}^{-}$ & 0.47 & 0.39 & $\cdots$ & $\cdots$ & ${<}0.01$ & 2.31 & 0.68 & 0.36 & $\cdots$ & 0.41 & 4.62 \\
			\hline
			$P_c^{\Delta}(4781)\frac{3}{2}^{-}$ & $\cdots$ & $\cdots$ & 0.87 & ${<}0.01$ & $\cdots$ & $\cdots$ & $\cdots$ & 0.32 & 1.18 & 1.44 & 3.81 \\
			\hline
			$P_c^{\Delta}(4765)\frac{1}{2}^{-}$ & $\cdots$ & $\cdots$ & $\cdots$ & 1.24 & $\cdots$ & $\cdots$ & 1.14 & 1.98 & $\cdots$ & 0.05 & 4.41 \\
			$P_c^{\Delta}(4799)\frac{1}{2}^{-}$ & $\cdots$& $\cdots$ & $\cdots$ & 1.17 & $\cdots$ & $\cdots$ & 0.12 & 0.93 & $\cdots$ & 4.74 & 6.96 \\
			\hline\hline
		\end{tabular}
	\end{table*}
	
\begin{table*}[hptb]
	\centering
\caption{Predicted masses (MeV), binding energies (B.E.) (MeV), OBE potential contributions $\langle V_{\chi, \sigma, v} \rangle$ (MeV), and inter-cluster RMS radii $\sqrt{\langle r^{2}\rangle}$ (fm) for the
$\Lambda_c\bar{D}^{(*)}$ and $\Sigma_c^{(*)}\bar{D}^{(*)}$ molecules. The central values are obtained with the $\sigma$-meson mass $m_{\sigma}=458~\mathrm{MeV}$, which is the central value recently determined by the partial-wave analysis of the $\pi\pi$ scattering~\cite{Danilkin:2020pak}. The uncertainties of the binding energies arising from the large uncertainty of the mass of $\sigma$ meson $m_{\sigma}=458^{+92}_{-58}~\mathrm{MeV}$ are also incorporated, which are denoted by the superscript and subscript. }
	\label{tab:molecular-spectrum-msigma-shift}
	\begin{tabular}{@{}cccccccccccc@{}}
		\hline
		{Hadron pair} & {$I(J^{P})$} & {Mass} & {Threshold} & {B.E.}
		& {$-\langle V_{\pi}\rangle$} & {$-\langle V_{\sigma}\rangle$}
		& {$-\langle V_{\eta}\rangle$} & {$-\langle V_{\eta^{\prime}}\rangle$}
		& {$-\langle V_{\rho}\rangle$}
		& {$-\langle V_{\omega}\rangle$}
		& {$\sqrt{\langle r^{2}\rangle}$} \\
		\hline
		
		\multicolumn{12}{l}{\textit{$I=1/2$ $\Lambda_c\bar{D}^{(*)}$ sector}} \\
		\hline
		$\Lambda_{c}\bar{D}$
		& $\frac{1}{2}(\frac{1}{2}^{-})$
		& $4150.44^{+0.31}_{-0.50}$ & 4151.00 & $0.57^{-0.32}_{+0.49}$
		& $\cdot\cdot\cdot$ & $0.78^{-0.38}_{+0.58}$ & $\cdot\cdot\cdot$ &$\cdot\cdot\cdot$
		& $\cdot\cdot\cdot$ & $-0.21^{+0.06}_{-0.09}$ & $5.02^{+0.12}_{-0.16}$ \\
		\hline
		
		\multirow{2}{*}{$\Lambda_{c}\bar{D}^{*}$}
		& $\frac{1}{2}(\frac{1}{2}^{-})$
		& $4292.44^{+0.31}_{-0.48}$ & \multirow{2}{*}{4293.00} & $0.56^{-0.31}_{+0.48}$
		& $\cdot\cdot\cdot$ & $0.77^{-0.37}_{+0.56}$ & $\cdot\cdot\cdot$ & $\cdot\cdot\cdot$
		& $\cdot\cdot\cdot$ & $-0.21^{+0.06}_{-0.09}$ & $5.03^{+0.12}_{-0.16}$ \\
		& $\frac{1}{2}(\frac{3}{2}^{-})$
		& $4292.44^{+0.31}_{-0.48}$ &  & $0.56^{-0.31}_{+0.48}$
		& $\cdot\cdot\cdot$ & $0.77^{-0.37}_{+0.56}$ & $\cdot\cdot\cdot$ & $\cdot\cdot\cdot$
		& $\cdot\cdot\cdot$ & $-0.21^{+0.06}_{-0.09}$ & $5.03^{+0.12}_{-0.16}$ \\
		\hline
		
		\multicolumn{12}{l}{\textit{$I=1/2$ $\Sigma_c^{(*)}\bar{D}^{(*)}$ sector}} \\
		\hline
		$\Sigma_{c}\bar{D}$
		& $\frac{1}{2}(\frac{1}{2}^{-})$
		& $4318.10^{+3.09}_{-4.28}$ & 4324.00 & $5.90^{-3.09}_{+4.28}$
		& $\cdot\cdot\cdot$ & $4.41^{-2.46}_{+3.53}$ & $\cdot\cdot\cdot$ & $\cdot\cdot\cdot$
		& $2.90^{-1.23}_{+1.48}$ & $-1.41^{+0.60}_{-0.72}$ & $3.79^{+0.58}_{-0.62}$ \\
		\hline
		
		$\Sigma_{c}^{*}\bar{D}$
		& $\frac{1}{2}(\frac{3}{2}^{-})$
		& $4382.35^{+2.95}_{-4.12}$ & 4388.00 & $5.65^{-2.95}_{+4.12}$
		& $\cdot\cdot\cdot$ & $4.23^{-2.35}_{+3.39}$ & $\cdot\cdot\cdot$ & $\cdot\cdot\cdot$
		& $2.77^{-1.16}_{+1.41}$ & $-1.35^{+0.57}_{-0.69}$ & $3.83^{+0.56}_{-0.61}$ \\
		\hline
		
		\multirow{2}{*}{$\Sigma_{c}\bar{D}^{*}$}
		& $\frac{1}{2}(\frac{1}{2}^{-})$
		& $4437.42^{+8.02}_{-6.54}$ & \multirow{2}{*}{4461.00} & $23.58^{-8.01}_{+6.54}$
		& $6.15^{-1.60}_{+1.07}$ & $12.94^{-5.30}_{+4.75}$
		& $-0.28^{+0.08}_{-0.07}$ & $-0.01^{+0.01}_{-0.00}$
		& $9.27^{-2.34}_{+1.54}$ & $-4.49^{+1.13}_{-0.74}$ & $2.02^{+0.60}_{-0.30}$ \\
		& $\frac{1}{2}(\frac{3}{2}^{-})$
		& $4458.31^{+1.34}_{-2.09}$ &  & $2.69^{-1.34}_{+2.10}$
		& $-0.35^{+0.16}_{-0.23}$ & $2.27^{-1.20}_{+1.92}$
		& $\sim 0$ & $\sim 0$
		& $1.49^{-0.56}_{+0.78}$ & $-0.73^{+0.27}_{-0.37}$ & $4.39^{+0.33}_{-0.42}$ \\
		\hline
		
		\multirow{3}{*}{$\Sigma_{c}^{*}\bar{D}^{*}$}
		& $\frac{1}{2}(\frac{1}{2}^{-})$
		& $4496.09^{+8.09}_{-6.52}$ & \multirow{3}{*}{4525.00} & $28.91^{-8.09}_{+6.52}$
		& $9.02^{-1.78}_{+1.18}$ & $14.74^{-5.34}_{+4.71}$
		& $-0.43^{+0.10}_{-0.08}$ & $-0.02^{+0.01}_{-0.01}$
		& $10.84^{-2.10}_{+1.38}$ & $-5.24^{+1.02}_{-0.67}$ & $1.76^{+0.43}_{-0.20}$ \\
		& $\frac{1}{2}(\frac{3}{2}^{-})$
		& $4513.40^{+5.60}_{-5.85}$ &  & $11.60^{-5.60}_{+5.85}$
		& $1.58^{-0.66}_{+0.59}$ & $7.50^{-3.96}_{+4.40}$
		& $-0.06^{+0.03}_{-0.03}$ & $0.01^{+0.01}_{-0.01}$
		& $5.01^{-1.97}_{+1.72}$ & $-2.43^{+0.96}_{-0.84}$ & $3.02^{+0.78}_{-0.58}$ \\
		& $\frac{1}{2}(\frac{5}{2}^{-})$
		& $4523.01^{+0.95}_{-1.48}$ &  & $1.99^{-0.95}_{+1.48}$
		& $-0.36^{+0.16}_{-0.23}$ & $1.75^{-0.90}_{+1.43}$
		& $\sim0$ & $\sim0$
		& $1.16^{-0.40}_{+0.56}$ & $-0.57^{+0.20}_{-0.27}$ & $4.55^{+0.27}_{-0.33}$ \\
		\hline
		
		\multicolumn{12}{l}{\textit{$I=3/2$ $\Sigma_c^{(*)}\bar{D}^{(*)}$ sector}} \\
		\hline
		$\Sigma_{c}\bar{D}$
		& $\frac{3}{2}(\frac{1}{2}^{-})$
		& $4323.77^{+0.16}_{-0.22}$ & 4324.00 & $0.23^{-0.16}_{+0.22}$
		& $\cdot\cdot\cdot$ & $0.46^{-0.21}_{+0.29}$ & $\cdot\cdot\cdot$ & $\cdot\cdot\cdot$
		& $-0.12^{+0.03}_{-0.04}$ & $-0.11^{+0.02}_{-0.04}$ & $5.18^{+0.07}_{-0.09}$ \\
		\hline
		
		$\Sigma_{c}^{*}\bar{D}$
		& $\frac{3}{2}(\frac{3}{2}^{-})$
		& $4387.77^{+0.16}_{-0.22}$ & 4388.00 & $0.23^{-0.16}_{+0.22}$
		& $\cdot\cdot\cdot$ & $0.46^{-0.21}_{+0.29}$ & $\cdot\cdot\cdot$ &$\cdot\cdot\cdot$
		& $-0.12^{+0.03}_{-0.03}$ & $-0.11^{+0.02}_{-0.04}$ & $5.18^{+0.07}_{-0.09}$ \\
		\hline
		
		\multirow{2}{*}{$\Sigma_{c}\bar{D}^{*}$}
		& $\frac{3}{2}(\frac{1}{2}^{-})$
		& $4460.71^{+0.15}_{-0.21}$ & \multirow{2}{*}{4461.00} & $0.29^{-0.15}_{+0.21}$
		& $\sim 0$ & $0.44^{-0.19}_{+0.27}$
		& $\sim0$ & $\sim0$
		& $-0.08^{+0.02}_{-0.02}$ & $-0.08^{+0.02}_{-0.02}$ & $5.15^{+0.07}_{-0.08}$ \\
		& $\frac{3}{2}(\frac{3}{2}^{-})$
		& $4460.79^{+0.16}_{-0.23}$ &  & $0.21^{-0.16}_{+0.24}$
		& $0.01^{-0.01}_{+0.01}$ & $0.48^{-0.22}_{+0.31}$
		& $\sim 0$ & $\sim 0$
		& $-0.14^{+0.03}_{-0.04}$ & $-0.14^{+0.04}_{-0.04}$ & $5.19^{+0.07}_{-0.09}$ \\
		\hline
		
		\multirow{3}{*}{$\Sigma_{c}^{*}\bar{D}^{*}$}
		& $\frac{3}{2}(\frac{1}{2}^{-})$
		& $4524.69^{+0.15}_{-0.21}$ & \multirow{3}{*}{4525.00} & $0.31^{-0.15}_{+0.21}$
		& $\sim0$ & $0.44^{-0.19}_{+0.27}$
		& $\sim0$ & $\sim 0$
		& $-0.07^{+0.01}_{-0.02}$ & $-0.07^{+0.02}_{-0.02}$ & $5.15^{+0.07}_{-0.09}$ \\
		& $\frac{3}{2}(\frac{3}{2}^{-})$
		& $4524.75^{+0.15}_{-0.22}$ &  & $0.25^{-0.15}_{+0.22}$
		& $\sim 0$ & $0.45^{-0.20}_{+0.27}$
		& $\sim 0$ & $\sim 0$
		& $-0.10^{+0.02}_{-0.03}$ & $-0.10^{+0.03}_{-0.02}$ & $5.17^{+0.07}_{-0.09}$ \\
		& $\frac{3}{2}(\frac{5}{2}^{-})$
		& $4524.79^{+0.17}_{-0.24}$ &  & $0.21^{-0.17}_{+0.24}$
		& $0.02^{-0.01}_{+0.02}$ & $0.49^{-0.23}_{+0.32}$
		& $\sim 0$ & $\sim 0$
		& $-0.15^{+0.04}_{-0.05}$ & $-0.15^{+0.04}_{-0.05}$ & $5.19^{+0.08}_{-0.09}$ \\
		\hline\hline
	\end{tabular}
\end{table*}

	\begin{table*}[hptb]
		\centering
		\caption[Fall-apart decay widths of molecules.]{
			Partial fall-apart decay widths $\Gamma_i$ (keV) of the
			$\Lambda_c\bar{D}^{(*)}$ and $\Sigma_c^{(*)}\bar{D}^{(*)}$
		    molecules with $I=1/2$ and $I=3/2$.}
		\label{tab:bound-state-A}
		\begin{tabular}{lccccccccccc}
			\hline\hline
			State & Mass (MeV)
			& ~~$\Gamma_{p\eta_{c}}$~~ &~~ $\Gamma_{p J/\psi}$~~
			& ~~$\Gamma_{\Delta\eta_{c}}$~~ & ~~$\Gamma_{\Delta J/\psi}$~~
			& ~~$\Gamma_{\Lambda_{c}\bar{D}^{0}}$~~ & ~~$\Gamma_{\Lambda_{c}\bar{D}^{*0}}$~~
			& ~~$\Gamma_{\Sigma_{c}\bar{D}}$~~ &~~ $\Gamma_{\Sigma_{c}\bar{D}^{*}}$~~
			& ~~$\Gamma_{\Sigma_{c}^{*}\bar{D}}$~~
			& $\Gamma_{sum}$ \\
			\hline
			
			$[\Lambda_{c}\bar{D}]_{1/2^-}^{1/2}$
			& 4150.44
			& 10.38 & 25.18
			& $\cdot\cdot\cdot$ & $\cdot\cdot\cdot$
			& $\cdots$ & $\cdots$
			& $\cdots$ & $\cdots$
			& $\cdots$
			& 35.55 \\
			\hline
			
			$[\Lambda_{c}\bar{D}^{*}]_{1/2^-}^{1/2}$
			& 4292.44
			& 8.86 & 2.30
			& $\cdots$ & $\cdots$
			& 4.29 & $\cdots$
			& $\cdots$ & $\cdots$
			& $\cdots$
			& 15.46 \\
			
			$[\Lambda_{c}\bar{D}^{*}]_{3/2^-}^{1/2}$
			& 4292.44
			& $\cdots$ & 25.14
			& $\cdots$ & $\cdots$
			& $\cdots$ & $\cdots$
			& $\cdots$ & $\cdots$
			& $\cdots$
			& 25.14 \\
			\hline
			
			$[\Sigma_{c}\bar{D}]_{1/2^-}^{1/2}$
			& 4318.10
			& 127.46 & 41.06
			& $\cdots$ & $\cdots$
			& 2.91 & 231.02
			& $\cdots$ & $\cdots$
			& $\cdots$
			& 402.45 \\
			\hline
			
			$[\Sigma_{c}^{*}\bar{D}]_{3/2^-}^{1/2}$
			& 4382.35
			& $\cdots$ & 106.13
			& $\cdots$ & $\cdots$
			& $\cdots$ & 236.22
			& $\cdots$ & $\cdots$
			& $\cdots$
			& 342.35 \\
			\hline
			
			$[\Sigma_{c}\bar{D}^{*}]_{1/2^-}^{1/2}$
			& 4437.42
			& 53.20 & 602.24
			& $\cdots$ & $\cdots$
			& 13.52 & 200.98
			& 17.52 & $\cdots$
			& $\cdots$
			& 887.47 \\
			
			$[\Sigma_{c}\bar{D}^{*}]_{3/2^-}^{1/2}$
			& 4458.31
			& $\cdots$ & 18.84
			& $\cdots$ & $\cdots$
			& $\cdots$ & 3.09
			& $\cdots$ & $\cdots$
			& 4.96
			& 26.88 \\
			\hline
			
			$[\Sigma_{c}^{*}\bar{D}^{*}]_{1/2^-}^{1/2}$
			& 4496.09
			& 133.48 & 46.31
			& $\cdots$ & $\cdots$
			& 20.95 & 0.03
			& 0.57 & 1.99
			& $\cdots$
			& 203.33 \\
			
			$[\Sigma_{c}^{*}\bar{D}^{*}]_{3/2^-}^{1/2}$
			& 4513.40
			& $\cdots$ & 105.54
			& $\cdots$ & $\cdots$
			& $\cdots$ & 23.04
			& $\cdots$ & 38.75
			& 4.93
			& 172.26 \\
			
			$[\Sigma_{c}^{*}\bar{D}^{*}]_{5/2^-}^{1/2}$
			& 4523.01
			& $\cdots$ & $\cdots$
			& $\cdots$ & $\cdots$
			& $\cdots$ & $\cdots$
			& $\cdots$ & $\cdots$
			& $\cdots$
			& $\cdots$ \\
			\hline
			
			$[\Sigma_{c}\bar{D}]_{1/2^-}^{3/2}$
			& 4323.77
			& $\cdots$ & $\cdots$
			& $\cdots$ & $\cdots$
			& $\cdots$ & $\cdots$
			& $\cdots$ & $\cdots$
			& $\cdots$
			& $\cdots$ \\
			\hline
			
			$[\Sigma_{c}^{*}\bar{D}]_{3/2^-}^{3/2}$
			& 4387.77
			& $\cdots$ & $\cdots$
			& 14.16 & 21.02
			& $\cdots$ & $\cdots$
			& $\cdots$ & $\cdots$
			& $\cdots$
			& 35.18 \\
			\hline
			
			$[\Sigma_{c}\bar{D}^{*}]_{1/2^-}^{3/2}$
			& 4460.71
			& $\cdots$ & $\cdots$
			& $\cdots$ & 7.62
			& $\cdots$ & $\cdots$
			& 0.10 & $\cdots$
			& $\cdots$
			& 7.72 \\
			
			$[\Sigma_{c}\bar{D}^{*}]_{3/2^-}^{3/2}$
			& 4460.79
			& $\cdots$ & $\cdots$
			& 14.79 & 32.60
			& $\cdots$ & $\cdots$
			& $\cdots$ & $\cdots$
			& 0.54
			& 47.94 \\
			\hline
			
			$[\Sigma_{c}^{*}\bar{D}^{*}]_{1/2^-}^{3/2}$
			& 4524.69
			& $\cdots$ & $\cdots$
			& $\cdots$ & 0.99
			& $\cdots$ & $\cdots$
			& 0.36 & 0.98
			& $\cdots$
			& 2.33 \\
			
			$[\Sigma_{c}^{*}\bar{D}^{*}]_{3/2^-}^{3/2}$
			& 4524.75
			& $\cdots$ & $\cdots$
			& 6.39 & 1.36
			& $\cdots$& $\cdots$
			& $\cdots$ & 3.86
			& 0.34
			& 11.95 \\
			
			$[\Sigma_{c}^{*}\bar{D}^{*}]_{5/2^-}^{3/2}$
			& 4524.79
			& $\cdots$ & $\cdots$
			& $\cdots$ & 59.07
			& $\cdots$ & $\cdots$
			& $\cdots$ & $\cdots$
			& $\cdots$
			& 59.07 \\
			\hline\hline
		\end{tabular}
	\end{table*}

\begin{figure}[htbp]
 \centering \epsfxsize=7.8 cm \epsfbox{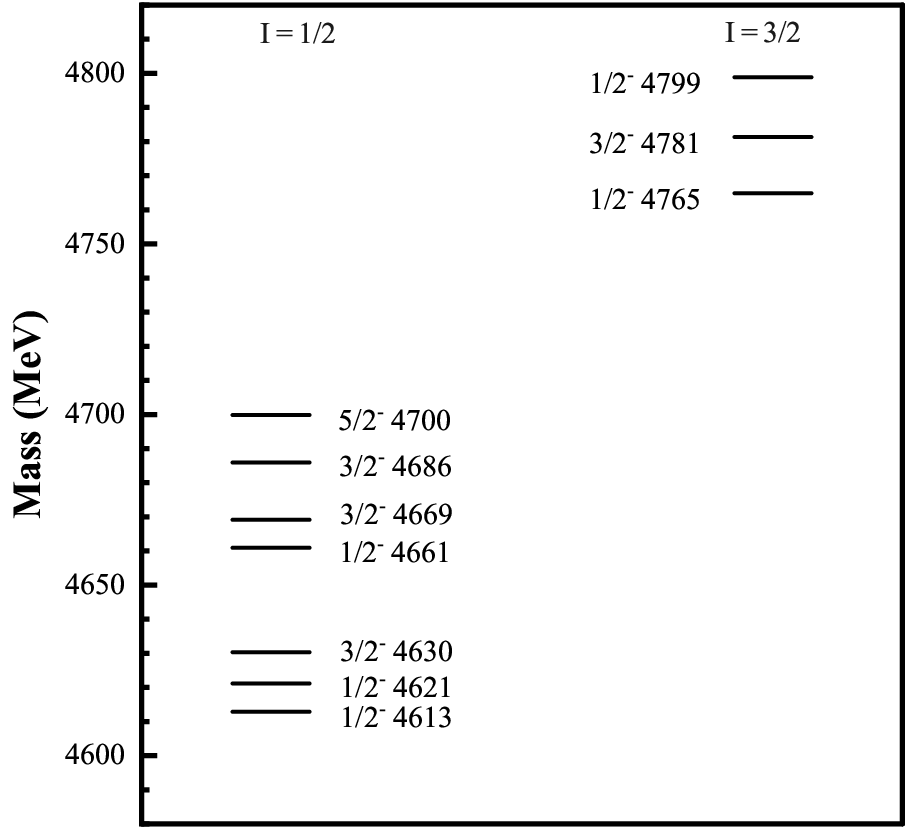}
 \caption{Mass spectrum of the $1S$-wave compact pentaquark states composed of $nnnc\bar{c}$.
}\label{fig:mass_compact}
\end{figure}

\begin{figure*}[htbp]
 \centering \epsfxsize=12.6 cm \epsfbox{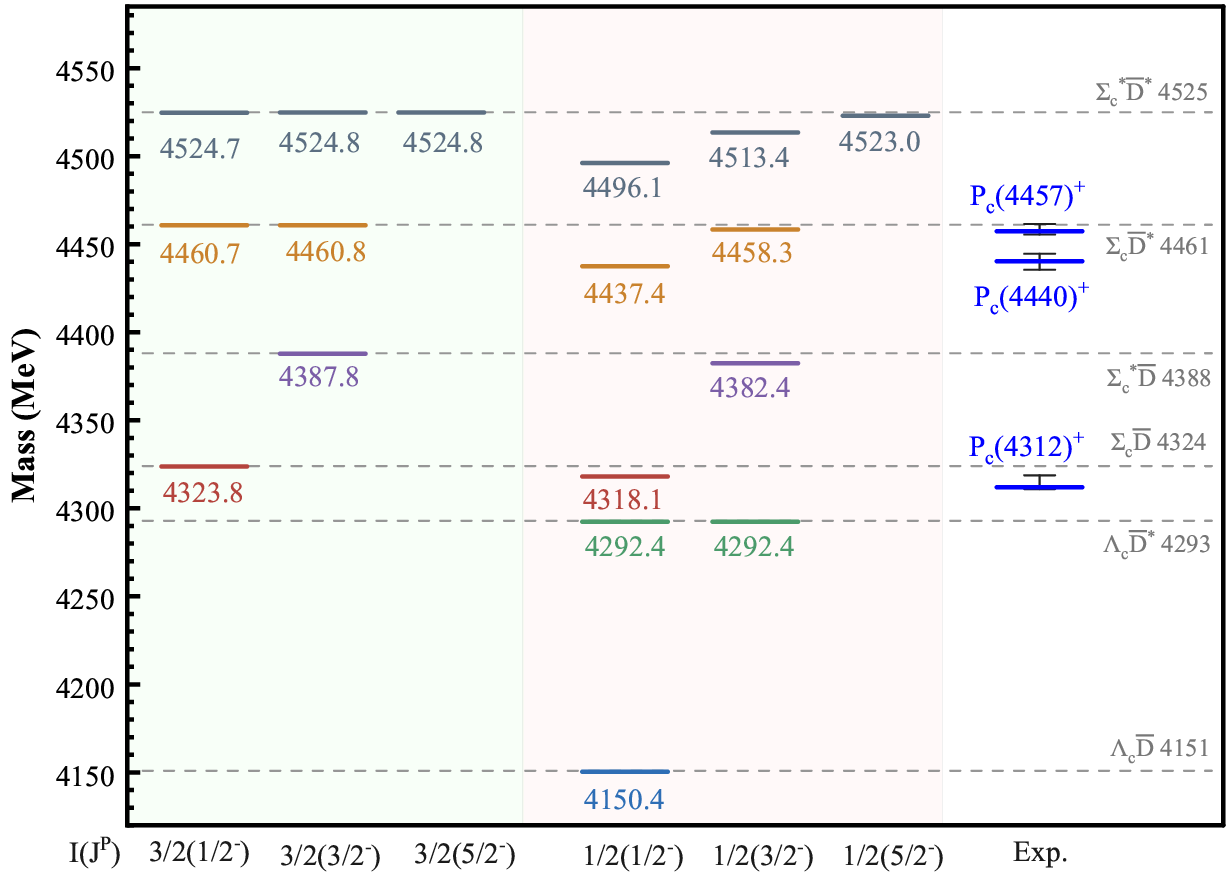}
 \caption{Mass spectrum of the $1S$-wave $\Sigma_c^{(*)}\bar{D}^{(*)}$ and $\Lambda_c^{(*)}\bar{D}^{(*)}$ molecules.
 The data of the $P_c(4312)^+$, $P_c(4440)^+$, and $P_c(4457)^+$ are taken from the PDG~\cite{ParticleDataGroup:2024cfk}.
}\label{fig:mass_Mole}
\end{figure*}

\begin{figure*}[htbp]
 \centering \epsfxsize=12.6 cm \epsfbox{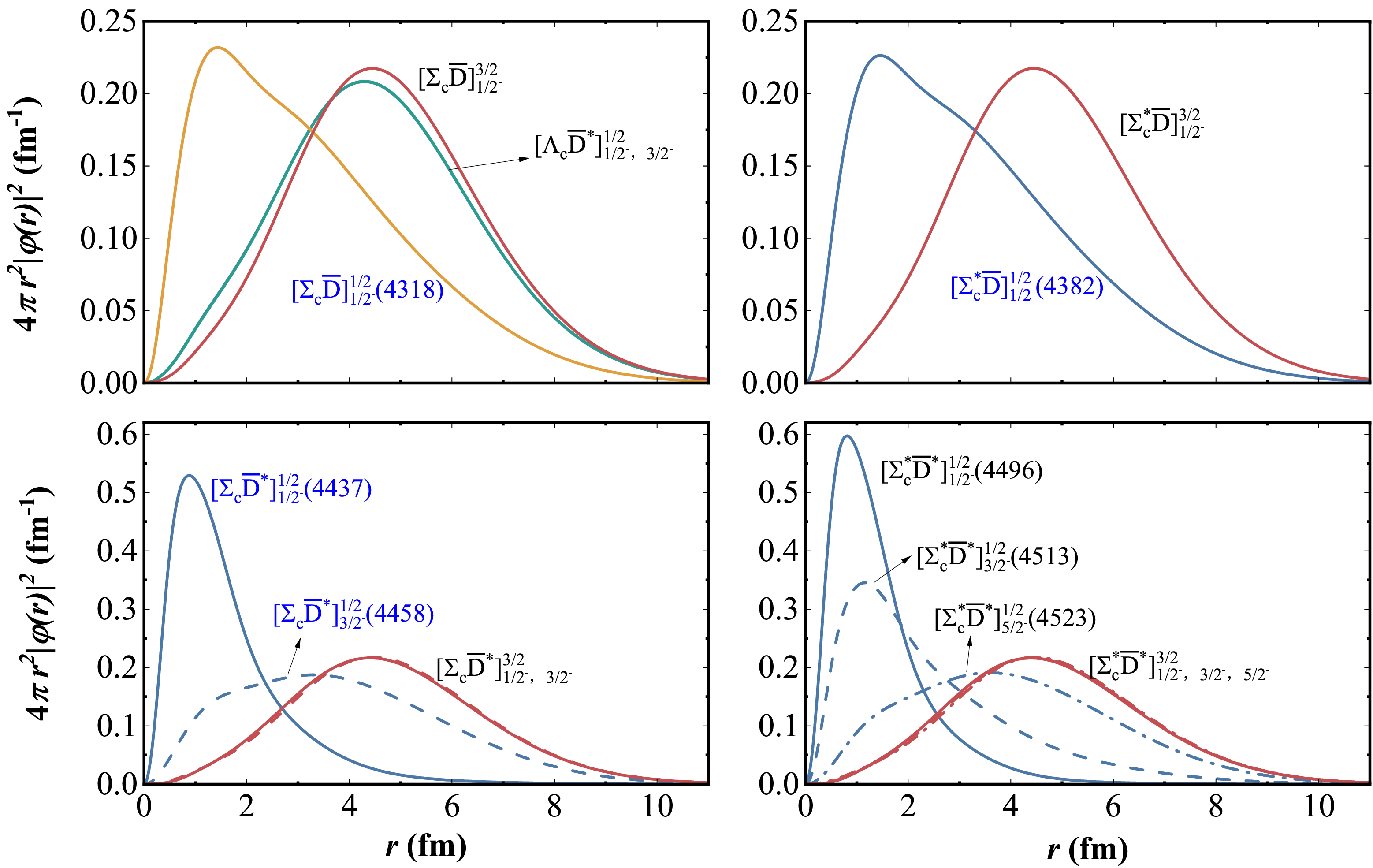}
 \caption{Predicted probability density $|\varphi(r)|^2$ for the $1S$-wave $\Sigma_c^{(*)}\bar{D}^{(*)}$ and $\Lambda_c^{(*)}\bar{D}^{(*)}$ molecules. $r$ stands for the distance between the centers of mass of two hadrons in a molecule.
}\label{fig:Mole}
\end{figure*}

\section{RESULTS AND DISCUSSIONS}\label{sec:results}
	
In this section, we present the numerical results for both the compact and molecular-type $qqqc\bar{c}$ pentaquark states.
The mass spectra, and fall-apart decay properties, and possible assignments of the observed $P_{c}$ states
are discussed in detail.

\subsection{Compact pentaquark states}\label{sec:compact}
	

Taking the configurations with the same $IJ^P$ numbers listed in Table~\ref{tab:configurations}
as basis vectors, and diagonalizing Hamiltonian matrix, one can obtain the masses and wave functions for the pentaquark states.
It should be mentioned that the $\mathbf{1}_c \otimes \mathbf{1}_c$ configuration cannot
form a bound system and also do nearly not mix with the $\mathbf{8}_c \otimes \mathbf {8}_c$ configurations
due to the absence of the OBE and OGE interactions between $(qqq)_{\mathbf{1}_c}$
and $(c\bar{c})_{\mathbf{1}_c}$. Thus, the pentaquark states that we obtain in the present work
are contributed by the $\mathbf{8}_c \otimes \mathbf{8}_c$ configurations.
The detailed results are given in Table~\ref{tab:mass-spectrum}.
For clarity, the obtained mass spectrum is plotted in Fig.~\ref{fig:mass_compact}.

For the isospin $I=1/2$ states, there are seven states which lie in a narrow mass range of $\sim 4610-4700$ MeV.
While the three $I=3/2$ states lie in the mass range of $\sim 4760-4800$ MeV,
which is about 100 MeV above that of the $I=1/2$ sector. These obtained pentaquark states
should be compact structures, since the root-mean-square radii between any two quarks
are predicted to be in the range of $0.63-0.88$ fm (see Table~\ref{tab:hamiltonian}).
Furthermore, from Table~\ref{tab:mass-spectrum}, it is seen that in the isospin $I=1/2$ states with $J^P=1/2$ and $3/2$,
there is significant configuration mixing. For example, the low-lying state $P_c^N(4630)\frac{3}{2}^-$
as a $1S_{1/2(3/2^-)_4}$ ($\sim86\%$) dominant state has a significant component of $1S_{1/2(3/2^-)_1}$
($\sim14\%$).

To see the dynamical roles of each part of the Hamiltonian,
we calculate their average values, which are listed in Table~\ref{tab:hamiltonian}.
It is found that in the Hamiltonian, the kinetic term $\langle T\rangle$, linear confinement potential $\langle V^{\mathrm{Conf}}\rangle$, and color-Coulomb potential $\langle V^{\mathrm{Coul}}\rangle$ are three main parts, they govern the mass of a pentaquark state.
Neglecting the other contributions, the pentaquark states have a nearly degenerate mass
of $\langle T\rangle+\langle V^{\mathrm{conf}}\rangle+\langle V^{\mathrm{Coul}}\rangle\simeq 4705$ MeV.
Their mass splitting is governed by the spin-dependent perturbation terms: the chromomagnetic part arising from one gluon exchange
and the OBE parts contributed by $\pi$, $\omega$, and $\rho$ exchanges. For example, for the
$I=3/2$ states, the $\omega$- and $\rho$-exchange terms contribute a sizeable repulsive
potential of $\langle V^{\rho}+V^{\omega}\rangle\sim 50$ MeV, which leads to a significant
gap between the $I=1/2$ and $I=3/2$ states. The $\sigma$ exchange always contributes a
sizable attractive potential $\langle V^{\sigma}\rangle\sim (-50,-40)$ MeV for all of the pentaquark states.
Thus, the role of $\sigma$ exchange in the compact states is easily hidden by the parameters
of zero energy and constituent quark mass.
For the $I=1/2$ states, due to a significant cancelation among the $\omega$- and $\rho$-exchange terms,
the total contributions from these vector meson exchanges are negligibly small.
The contributions from the $\eta$ and $\eta'$ exchanges are negligibly small as well duo to the weak couplings.


All of the obtained compact $qqqc\bar{c}$ states lie far above the mass threshold of $J/\psi p$ ($\sim4035$ MeV),
thus, they can dissociate into some low-lying baryon-meson pairs through the quark rearrangement mechanism.
By using the obtained wave functions, we further evaluate the fall-apart decay properties of the compact $qqqc\bar{c}$ states.
Our results are collected in Table~\ref{tab:compact-pentaquark-new}. It is found that the compact $qqqc\bar{c}$ states
have a very narrow fall-apart decay width of serval MeV.
The $I=1/2$ states $P_c^N(4630,4669,4686)\frac{3}{2}^-$ and $P_c^N(4613,4621,4661)\frac{1}{2}^-$
may have a large decay rate into the $J/\psi p$ channel, its partial width accounts for
$\sim10-20\%$ of the total fall-apart width. These narrow and compact $qqqc\bar{c}$ states may
have potentials to be observed in the $J/\psi p$ final state.
	
Finally, it should be mentioned that the $P_c(4312)^+$, $P_c(4440)^+$, and $P_c(4457)^+$ observed at LHCb~\cite{LHCb:2019kea} cannot
be assigned to any compact pentaquark states composed of $qqqc\bar{c}$, whose masses are predicted to be
$\sim200-300$ MeV larger than the observations.

\subsection{Hadronic molecules}\label{sec:molecular}

The discrepancies between the compact pentaquark spectrum and experimental observations suggest that further crucial mechanisms different from the color exchange potential for the compact pentaquark states should be present. For states, which have strong $S$-wave coupling to the nearby open threshold, it interactions can play a crucial role in shaping the hadron spectroscopy. In other words, the hadronic molecule component can be part of the wave function of the initial hadron, and even becomes dominant if the near-threshold interaction is strong enough. In this Subsection we investigate hadronic molecule scenario at quark level, where two color-singlet clusters will be the effective degrees of freedom to be considered.
First, we calculate the mass spectra for the $1S$-wave $\Sigma_c^{(*)}\bar{D}^{(*)}$
and $\Lambda_c\bar{D}^{(*)}$ molecules as given in Table~\ref{tab:molecular_states}.
To reduce the model dependency of the predicted masses for the molecules, we do not adopt the theoretical masses
obtained from Eq.~(\ref{Massa}) directly. Using the molecular wave functions obtained from this equation,
we first calculate the binding energies with
\begin{eqnarray}
B.E.=-\sum_{i<j,i\in H_1, j\in H_2} \left\langle [H_1H_2]^{II_z}_{JJ_z} \left|V^{OBE}_{ij}\right| [H_1H_2]^{II_z}_{JJ_z} \right\rangle.
\end{eqnarray}
Then, the mass for the molecular state $[H_1H_2]^{II_z}_{JJ_z}$ is determined by
\begin{eqnarray}
M_{mole}=M_{H_1}+M_{H_2}-B.E.,
\end{eqnarray}
where we adopt the measured masses $M_{H_1}$ and $M_{H_2}$ for hadrons $H_1$ and $H_2$, respectively.
Our results are summarized in Table~\ref{tab:molecular-spectrum-msigma-shift}. The obtained mass spectrum
is also plotted in Fig.~\ref{fig:mass_Mole} for clarity. It should be mentioned that
the mixing between different molecular configurations with the same $IJ^P$ numbers is negligibly small, which is
not explicitly given for simplicity. The obtained hadronic molecules are loosely bound states of two hadrons,
which can be obviously seen from the spatial wave functions as shown in Fig.~\ref{fig:Mole} and their RMS radii given in Table~\ref{tab:molecular-spectrum-msigma-shift}. Additionally, by using the obtained masses and wave functions for the molecules, the fall-apart decay properties
are further evaluated, the results are collected in Table~\ref{tab:bound-state-A}.

\subsubsection{$\Sigma_c\bar{D}$ system and $P_c(4312)$}

From Table~\ref{tab:molecular-spectrum-msigma-shift}, one can see that the $\Sigma_c\bar{D}$ can form a bound state with quantum numbers $IJ^P=\frac{1}{2}\frac{1}{2}^-$ and a binding energy of $\sim6$ MeV (denoted by $[\Sigma_c\bar{D}]_{1/2^-}^{1/2}(4318)$) due to sizeable attractive interactions from the $\sigma$- and $\rho$-meson exchanges. The $[\Sigma_c\bar{D}]_{1/2^-}^{1/2}(4318)$ should
be a loose structure since the root-mean-square distance between $\Sigma_c$ and $\bar{D}$ is predicted to
be $\sim3.8$ fm. The $[\Sigma_c\bar{D}]_{1/2^-}^{1/2}(4318)$ may have significant decay
rates into $\eta_c p$, $J/\psi p$, and $\Lambda_c \bar{D}^*$ channels with ratios
\begin{eqnarray}
	\Gamma[J/\psi p]:\Gamma[\eta_c p]:\Gamma[\Lambda_c \bar{D}^*]\simeq 1.0:3.1:5.6.
\end{eqnarray}
However, the decay rate into the $\Lambda_c\bar{D}$ channel is negligibly small.

The $P_{c}(4312)^+$, which was observed by the LHCb collaboration in the $J/\psi\, p$
invariant mass spectrum, favors the molecular state assignment $[\Sigma_c\bar{D}]_{1/2^-}^{1/2}(4318)$.
This assignment was also suggested in the literature, e.g.~\cite{Wang:2025ecf,Liu:2019tjn,He:2019ify,Chen:2019asm,Xiao:2019aya,Wang:2019ato}.
Our analysis shows that the fall-apart decay width is only $\mathcal{O}(1\%)$ of
the observed width $\Gamma_{exp}=9.8\pm2.7^{+3.7}_{-4.5}$ MeV, which indicates there
are other dominant decay modes for the $P_{c}(4312)^+$ state.
Combining the predicted partial width of $\Gamma[J/\psi p]\simeq 0.04$ MeV with this measured total width,
we predict that
\begin{eqnarray}
Br[P_{c}(4312)^+\to J/\psi p]\simeq 4.1^{+11.2}_{-1.6}\times10^{-3},
\end{eqnarray}
which is consistent with the extracted range of $0.05\%<Br[P_{c}(4312)^+\to J/\psi p]<1.2\%$~\cite{Cao:2019kst} based on the
branching ratios and fractions measured by LHCb~\cite{LHCb:2015qvk} and
GlueX~\cite{GlueX:2019mkq} collaborations. To confirm the nature of $P_{c}(4312)^+$,
the other important decay modes $\eta_c p$ and $\Lambda_c\bar{D}^*$ are worth observing in future experiments.

The $\Sigma_c\bar{D}$ may form a very loosely bound state with quantum numbers $IJ^P=\frac{3}{2}\frac{1}{2}^-$
and a tiny binding energy of $\sim230$ keV due to a weak attractive interaction from the $\sigma$-meson exchange.
The root-mean-square distance between $\Sigma_c$ and $\bar{D}$ is predicted to
be a fairly large value of $\sim5.0$ fm. The $[\Sigma_c\bar{D}]_{1/2^-}^{3/2}$ may be a very narrow
state since its fall-apart decay modes are nearly forbidden.

\subsubsection{$\Sigma_c\bar{D}^*$ system and $P_c(4440,4457)$}

In the isospin $I=1/2$ sector, the $\Sigma_c\bar{D}^*$ can form two bound states with quantum numbers $IJ^P=\frac{1}{2}\frac{1}{2}^-$
and $IJ^P=\frac{1}{2}\frac{3}{2}^-$, i.e., $[\Sigma_c\bar{D}^*]_{1/2^-}^{1/2}(4437)$ and
$[\Sigma_c\bar{D}^*]_{3/2^-}^{1/2}(4458)$.

For the deep bound $[\Sigma_c\bar{D}^*]_{1/2^-}^{1/2}(4437)$,
the binding energy is predicted to be $\sim 24$ MeV, which is mainly contributed by the $\pi$-, $\sigma$-,
and $\rho$-meson exchanges. Due to their strong attractive integrations, the root-mean-square distance between $\Sigma_c$
and $\bar{D}^*$ is a relative small value, $2.0$ fm. The partial width ratio between the main fall-apart decay
channels $J/\psi p$ and $\Lambda_c\bar{D}^*$ is predicted to be
\begin{eqnarray}
\Gamma[J/\psi p]:\Gamma[\Lambda_c\bar{D}^*]\simeq 3:1.
\end{eqnarray}

For the $[\Sigma_c\bar{D}^*]_{3/2^-}^{1/2}(4458)$ state, the binding energy is shallow, i.e.,
$\mathrm{B.E}.\simeq 2.7$ MeV, which is mainly contributed by the $\sigma$-,
and $\rho$-meson exchanges, however, the $\pi$- and $\omega$-meson exchange
interactions exhibit weak repulsion. The $[\Sigma_c\bar{D}^*]_{3/2^-}^{1/2}(4458)$
is a very loose state with a fairly large root-mean-square radius of $\sim4.4$ fm, due
to the shallow binding energy.
While the $[\Sigma_c\bar{D}^*]_{3/2^-}^{1/2}(4458)$ state only weakly couples to the fall-apart decay
channels $J/\psi p$, $\Lambda_c\bar{D}^*$, and $\Sigma_c^*\bar{D}$.
The partial width of the dominant channel $J/\psi p$ is predicted to be $\sim 20$ keV.

The $P_{c}(4440)^+$ and $P_{c}(4457)^+$ observed at LHCb in the $J/\psi\, p$
invariant mass spectrum~\cite{LHCb:2019kea} can be assigned as the $\Sigma_c\bar{D}^*$ molecular states
$[\Sigma_c\bar{D}^*]_{1/2^-}^{1/2}(4437)$ and $[\Sigma_c\bar{D}^*]_{3/2^-}^{1/2}(4458)$, respectively.
Our assignments are consistent those of Refs.~\cite{Wang:2025ecf,Liu:2019tjn,He:2019ify,Chen:2019asm,Xiao:2019aya,Wang:2019ato}.
However, reversed spin-parity numbers for the $P_{c}(4440)^+$ and $P_{c}(4457)^+$ were also suggested
in the literature~\cite{Du:2021fmf,Yamaguchi:2019seo,Liu:2019zvb,Deng:2026gqe}. The measured widths of $P_{c}(4440)^+$ and $P_{c}(4457)^+$ are
$\Gamma_{exp}=20.6\pm4.9^{+8.7}_{-10.1}$ MeV and  $6.4\pm2.0^{+5.7}_{-1.9}$ MeV, respectively.
Combining these measured widths with our predicted partial widths for the $J/\psi p$ channel,
we predict that
\begin{eqnarray}
		Br[P_{c}(4440)^+\to J/\psi p]&\simeq & 2.9^{+7.8}_{-1.1}\%,\\
		Br[P_{c}(4457)^+\to J/\psi p]&\simeq & 2.9^{+4.6}_{-1.6}\times10^{-3},
\end{eqnarray}
which are consistent with the extracted ranges $3.55^{+1.43}_{-1.24}\times 10^{-3}<Br[P_{c}(4440)^+\to J/\psi p]<2\%$
and $1.70^{+0.77}_{-0.71}\times 10^{-3}<Br[P_{c}(4457)^+\to J/\psi p]<2\%$~\cite{Cao:2019kst} based on the
branching ratios and fractions measured by LHCb~\cite{LHCb:2015qvk} and
GlueX~\cite{GlueX:2019mkq} collaborations. The $P_{c}(4440)^+$ may have potential to be observed in
the $\Lambda_c\bar{D}^*$ channel as well, the branching fraction can reach up to
$Br[P_{c}(4440)^+\to\Lambda_c\bar{D}^*]\sim 1\%$.


Finally, it should be mentioned that in the $I=3/2$ sector there may exist two very loosely
bound states of $\Sigma_c\bar{D}^*$ with quantum numbers $J^P=1/2^-$ and $3/2^-$, respectively.
They are highly degenerate and lie about 200-300 keV below the mass threshold of $\Sigma_c\bar{D}^*$.
The weak attractive interaction between $\Sigma_c$ and $\bar{D}^*$ is contributed by the $\sigma$
exchange. In another word, the existence of the $\Sigma_c\bar{D}^*$ bound states with $IJ^P=\frac{3}{2}\frac{1}{2}^-$ and
$\frac{3}{2}\frac{3}{2}^-$ highly depend on the strength of the interaction from $\sigma$ exchange.
If they exist, they may very narrow states due to their very weak couplings to the main fall-apart
channels, such as $\eta_c\Delta$ and $J/\psi \Delta$.

\subsubsection{$\Sigma_c^*\bar{D}$ system and $P_c(4380)^+$ }

In the isospin $I=1/2$ sector, the $\Sigma_c^*\bar{D}$ can form a loosely bound state with quantum numbers
$J^P=\frac{3}{2}^-$, denoted by $[\Sigma_c^*\bar{D}]_{3/2^-}^{1/2}(4382)$, due to sizeable attractive
interactions from the $\sigma$- and $\rho$-meson exchanges. The binding energy is predicted to be
$B.E.\simeq 5.65$ MeV, which is comparable with that of $[\Sigma_c\bar{D}]_{1/2^-}^{1/2}(4318)$, thus they
have a similar root-mean-square radius $\sim 3.8$ fm. The $[\Sigma_c^*\bar{D}]_{3/2^-}^{1/2}(4382)$
may be a narrow state, whose fall-apart decays are dominated by the $J/\psi p$ and $\Lambda_c \bar{D}^*$
channels with a partial width ratio
\begin{eqnarray}
		\Gamma[J/\psi p]:\Gamma[\Lambda_c\bar{D}^*]\simeq 1.0:2.3.
\end{eqnarray}

The $[\Sigma_c^*\bar{D}]_{3/2^-}^{1/2}(4382)$ may have been observed in LHCb experiments.
The Run 1 data of the LHCb measurements in 2015 show a broad structure $P_c(4380)^+$ in the $J/\psi p$
invariant mass spectrum. The analysis of the combined data of Run 1 and 2 reported by LHCb in 2019 inidcates the $P_c(4380)^+$
should be a narrow state with a width of $\Gamma\simeq26$ MeV~\cite{Du:2019pij}. The $P_c(4380)^+$ structure may be signals of
$[\Sigma_c^*\bar{D}]_{3/2^-}^{1/2}(4382)$ predicted in theory. The Run 3 data of LHCb should help establish
the $[\Sigma_c^*\bar{D}]_{3/2^-}^{1/2}(4382)$ finally.

In the isospin $I=3/2$ sector, the $\Sigma_c^*\bar{D}$ may form a very
shallow bound state $[\Sigma_c^*\bar{D}]_{3/2^-}^{3/2}(4388)$ with a very
small binding energy of $B.E.\simeq 230$ keV due to the weak attractive interaction from the $\sigma$-meson exchange.
The $[\Sigma_c^*\bar{D}]_{3/2^-}^{3/2}(4388)$ as a very loosely bound state of $\Sigma_c^*\bar{D}$ has
a fairly large root-mean-square radius $\sim 5.2$ fm, and weakly couples to
the $\eta_c \Delta$ and $J/\psi \Delta$ channels with partial widths of a few tens keV.


\subsubsection{$\Sigma_c^*\bar{D}^*$ system}

In the isospin $I=1/2$ sector, the $\Sigma_c^*\bar{D}^*$ may form three bound states:
$[\Sigma_c^*\bar{D}^*]_{1/2^-}^{1/2}(4496)$, $[\Sigma_c^*\bar{D}^*]_{3/2^-}^{1/2}(4513)$, and $[\Sigma_c^*\bar{D}^*]_{5/2^-}^{1/2}(4523)$.
The $[\Sigma_c^*\bar{D}^*]_{1/2^-}^{1/2}(4496)$ is a deeply bound state
of $\Sigma_c^*\bar{D}^*$ with a fairly large binding energy of $B.E.\simeq 29$ MeV,
which is contributed by the $\sigma$-, $\rho$-, and $\pi$-meson exchanges. Its root-mean-square
radius $\sim 1.8$ fm is notably smaller than other states due to the strong
attractive interactions from $\sigma$-, $\rho$-, and $\pi$-meson exchanges.
The $[\Sigma_c^*\bar{D}^*]_{1/2^-}^{1/2}(4496)$ has a very narrow fall-apart width of $\mathcal{O}(200)$ keV,
which is nearly saturated by the $J/\psi p$ and $\eta_c p$ channels. The partial width ratio is predicted to be
\begin{eqnarray}
		\Gamma[J/\psi p]:\Gamma[\eta_c p]\simeq 1.0:2.9.
\end{eqnarray}
The $[\Sigma_c^*\bar{D}^*]_{1/2^-}^{1/2}(4496)$ may have potentials to be observed
in the $J/\psi p$ channel.

The $[\Sigma_c^*\bar{D}^*]_{3/2^-}^{1/2}(4513)$ has a moderate binding
energy of $B.E.\simeq 11$ MeV, which is mainly contributed by the $\sigma$-, $\rho$-meson
exchanges. Furthermore, the $\pi$-meson also provides a sizeable attractive interaction,
which is about $3-4$ times smaller that of $\sigma$ and $\rho$ mesons.
The $[\Sigma_c^*\bar{D}^*]_{3/2^-}^{1/2}(4513)$ has a very narrow fall-apart width of $\mathcal{O}(200)$ keV,
which is mainly contributed by the $J/\psi p$ and $\Sigma_c\bar{D}^*$ channels with a
partial width ratio
\begin{eqnarray}
\Gamma[J/\psi p]:\Gamma[\Sigma_c\bar{D}^*]\simeq 1.0:0.4.
\end{eqnarray}
To establish the $[\Sigma_c^*\bar{D}^*]_{1/2^-}^{1/2}(4496)$, the $J/\psi p$ channel is worth observing in future experiments.

The $[\Sigma_c^*\bar{D}^*]_{5/2^-}^{1/2}(4523)$ should have a very loose structure if it exist.
The root-mean-square radius reaches up to a fairly large value of $\sim 4.6$ fm due
to the weak binding energy, $B.E.\simeq 2.0$ MeV, which is about an order of magnitude
smaller than the other two low-lying bound states of $\Sigma_c^*\bar{D}^*$.
The weak attractive interactions between $\Sigma_c^*$ and $\bar{D}^*$
come from both $\sigma$ and $\rho$ exchanges.
The $[\Sigma_c^*\bar{D}^*]_{5/2^-}^{1/2}(4523)$ may be a very narrow state since
it has no fall-apart decay channels.

In the isospin $I=3/2$ sector, the existence of the $\Sigma_c^*\bar{D}^*$
bound states highly depends on the strength of the $\sigma$ exchange potential.
If they exist indeed, they should have very small binding energies of $B.E.\simeq 200-300$ keV
and fairly large root-mean-square radii of $\sim 5.2$ fm. They may be very narrow states since they weakly couples to
the main fall-apart decay channels.


\subsubsection{$\Lambda_c\bar{D}^{(*)}$ system}
	
The $\Lambda_c\bar{D}^{(*)}$ may form very loosely bound states with tiny binding energies
of $B.E.\simeq 500-600$ keV and large root-mean-square radii of $\sim 5.0$ fm.
The weak attractive interactions between $\Lambda_c$ and $\bar{D}^{(*)}$ arise
from the $\sigma$ exchange, while the $\omega$ exchange provides a tiny repulsive interaction.
It should be pointed out that the potential strength of the $\sigma$ exchange is not well
constrained by the experiment, when we reduce the coupling constant $g_{\sigma}$ a factor of $\sim2$,
the $\Lambda_c\bar{D}^{(*)}$ systems cannot form bound states. The possible existence of
a shallow $\Lambda_c\bar{D}$ bound state is also predicted
in Refs.~\cite{Chen:2017vai,Yan:2022wuz}, depending on the input model parameters.
If $\Lambda_c\bar{D}^{(*)}$ bound states exist indeed, they may have very narrow fall-apart width
of a few tens of keV. To confirm the existence of the $\Lambda_c\bar{D}^{(*)}$ states,
the $J/\psi p$ channel is worth observing in future experiments. More details about the spectrum
behavior and decays can be seen in Tables \ref{tab:molecular-spectrum-msigma-shift} and \ref{tab:bound-state-A}.

\section{Summary}\label{sec:summary}
	
In this work, we carry out a unified study of the low-lying $1S$-wave compact states and hadronic
molecules composed of $qqqc\bar{c}$ within a semirelativistic potential quark model. Then, we further evaluate the fall-apart decays by
combining the obtained spectra within the quark exchange model, since our obtained states lie far above the lowest
dissociation baryon-meson threshold. We expect our study provides useful information for
further exploring the hidden-charm pentaquarks in future experiments. The key points are summarized as follows.
	
For the compact pentaquarks, we predict seven states with isospin $I=1/2$, whose masses scatter in the range of $\sim4.6-4.7$ GeV.
While for the isospin $I=3/2$ sector, there are three states lying in the mass range of $\sim4.8$ GeV.
All of the obtained compact states have narrow fall-apart decay widths of a few MeV.
Some states with $I(J^P)=1/2(1/2^-)$ and $ 1/2(3/2^-)$ may have good potentials to be observed
in the $p J/\psi$ channel around the mass range of $4.6-4.7$ GeV.

The $\Sigma_c^{(*)}\bar{D}^{(*)}$ systems can form seven molecules with $I=1/2$.
Their binding energies scatter in the range of $\sim2-30$ MeV, which is mainly
contributed by the $\sigma$- and $\rho$-meson exchanges. In the $[\Sigma_c\bar{D}^*]_{1/2^-}^{1/2}(4437)$ and
$[\Sigma_c\bar{D}^*]_{3/2^-}^{1/2}(4458)$ states, the $\pi$-meson exchange also plays an important role.

The $\Sigma_c^{(*)}\bar{D}^{(*)}$ in the isospin $I=3/2$ sector and the $\Lambda_c\bar{D}^{(*)}$ may form very
loosely bound states due to the weak attractive interactions from $\sigma$ exchange,
which depends on the potential strength of the $\sigma$ exchange what we adopt.

The narrow states $P_c(4312)^+$, $P_{c}(4440)^+$, and $P_{c}(4457)^+$ observed at LHCb can
be assigned as the hadronic molecules $[\Sigma_c\bar{D}]_{1/2^-}^{1/2}(4318)$, $[\Sigma_c\bar{D}^*]_{1/2^-}^{1/2}(4437)$, and
$[\Sigma_c\bar{D}]_{3/2^-}^{1/2}(4458)$, respectively. The structure $P_c(4380)^+$ reported by LHCb in 2015 may contributed by the molecule $[\Sigma_c^*\bar{D}]_{3/2^-}^{1/2}(4382)$, which should be a narrow state other than a broad state reported by LHCb.
The $P_c(4312)^+$, $P_{c}(4440)^+$, and $P_c(4380)^+$ may have good potentials to be observed in the $\Lambda_c \bar{D}^*$
channel as well.


	\section*{Acknowledgements}
	
We thank Xiao-Hai Liu, Lu Meng, Ming-Sheng Liu, Fei Huang, and Rui Chen for useful discussions.
This work is supported by the National Natural Science Foundation of China under Grants Nos. 12235018 and 12175065.

	\bibliographystyle{unsrt}

\end{document}